\documentclass[3p,times]{elsarticle}

 \usepackage{multirow}
\usepackage[abs]{overpic}
\usepackage{xcolor}

\usepackage{pgfplotstable} 
\usepgfplotslibrary{statistics}
\usepackage{subfigure}
\usetikzlibrary{arrows,decorations.pathmorphing,backgrounds,positioning,fit,calc,3d,er,pgfplots.groupplots}
\usetikzlibrary{spy}
\usepgfplotslibrary{fillbetween}
\usetikzlibrary{patterns}
\pgfplotsset{compat=1.18,set layers}
\usepackage{booktabs}
\usepackage{comment}

\pgfplotsset{width=10cm,compat=1.9}
\pgfplotsset{compat=newest}

\usepackage[bold]{hhtensor}
\usepackage{float}
\usepackage{multirow}
\usepackage{nicefrac}
\usepackage{tikz}
\usetikzlibrary{positioning,fit,calc,decorations.pathreplacing}

\definecolor{diffcol}{RGB}{252,244,214}
\definecolor{inscol}{RGB}{221,238,255}
\definecolor{sinkcol}{RGB}{214,237,214}
\definecolor{reaccol}{RGB}{245,214,214}

\usepackage{cases}
\usetikzlibrary{arrows,decorations.pathmorphing,backgrounds,positioning,fit,calc,3d,er}
\usetikzlibrary{intersections,fillbetween}

\usepackage{todonotes}
\usepackage[most]{tcolorbox}
\usepackage{soul}

\usepackage{hyperref}
\hypersetup{
    colorlinks,%
    citecolor=blue,%
    filecolor=black,%
    linkcolor=magenta,%
    urlcolor=black
}

\usepackage{latexsym}
\usepgfplotslibrary{fillbetween}
\usepackage{lineno}
\usepackage{float}
\usepackage{booktabs} 
\pgfplotsset{compat=1.9}

\usepackage{mathtools}
\usepackage{comment}
\usepackage[dvipsnames]{xcolor}

\newcommand{\etal}{{\it et al.}}
\usepackage{tikz}
\usepackage{tikz-qtree}
\usetikzlibrary{shapes.geometric, arrows}
\usetikzlibrary{calc}
\tikzstyle{startstop} = [rectangle, rounded corners, minimum width=3cm, minimum height=1cm, text centered, draw=black, fill=white!30]
\tikzstyle{process} = [rectangle, minimum width=3cm, minimum height=1cm, text centered, draw=black, fill=white!30]
\tikzstyle{decision} = [diamond, minimum width=3cm, minimum height=1cm, text centered, draw=black, fill=white!30]
\tikzstyle{arrow} = [thick,->,>=stealth]
\usepackage{tkz-berge}
\usepackage{transparent}
\usepackage{sansmath}
\usepackage{bm}
\usetikzlibrary{shadings,intersections,fadings}
\usetikzlibrary{shapes.geometric, arrows.meta}
\usepackage{tikz-3dplot}
\usetikzlibrary{calc,3d,decorations.markings,backgrounds,positioning,intersections,shapes,quotes,angles}
\usepackage{nicefrac}
\usepackage{caption}
\usepackage{pdflscape}
\usepackage{graphicx}
\usepackage{caption,booktabs}
\usepackage{amsmath}
\usepackage{amstext} 

\usepackage[abs]{overpic}
\usepackage{subcaption}
\usepackage{array}
\usepackage{enumitem}

\usepackage[]{hyperref}
\hypersetup{
    colorlinks,%
    citecolor=blue,%
    filecolor=black,%
    linkcolor=blue,%
    urlcolor=black
}
\usepackage{cleveref}
\usepackage{tabularx}
\usepackage{multicol}

 \usepackage[bold]{hhtensor}
\usepackage{algorithm}
\usepackage{algpseudocode}

\biboptions{sort&compress}
 
\usepackage{scrextend}
 \usepackage{lineno}

\begin{document}

 \tdplotsetmaincoords{0}{0}
 
\begin{frontmatter} 

\title{Parameter-free prediction of irradiation defect structures in tungsten at room temperature using stochastic cluster dynamics}

\author[UCLA_MSE]{Sicong He}
\author[UCSD_MAE]{Brandon Schwendeman}
\author[UCSD_MAE]{George Tynan}
\author[UCLA_MSE,UCLA_MAE]{Jaime Marian}
\ead{jmarian@ucla.edu}
 
\address[UCLA_MSE]{Department of Materials Science and Engineering, University of California Los Angeles, Los Angeles, CA 90095, USA}
\address[UCSD_MAE]{Department of Mechanical and Aerospace Engineering, University of California San Diego, La Jolla, CA 92093, USA}
\address[UCLA_MAE]{Department of Mechanical and Aerospace Engineering, University of California Los Angeles, Los Angeles, CA 90095, USA}
 
\begin{abstract}
The foundations of irradiation damage theory were laid in the 1950s and 60s within the framework of chemical reaction kinetics. While helpful to analyze qualitative aspects of irradiation damage, the theory contained gaps that delayed its implementation and applicability as a predictive tool. The advent of computer simulations with atomistic resolution in the 80s and 90s revealed a series of mechanisms that have proved essential to understand key aspects of irradiation damage in crystalline solids. However, we still lack a comprehensive model that can connect atomic-level defect physics with experimental measurements of quantitative features of the irradiated microstructure. In this work, we present a mesoscale model that draws from our improved understanding of irradiation damage processes collected over the last few decades, bridging knowledge gained from our most sophisticated atomistic simulations with defect kinetics taking place over time scales many orders of magnitude larger than atomic interaction times. Importantly, the model contains no adjustable parameters and combines several essential pieces of irradiation damage physics: each playing an irreplaceable role in the context of the full model but of limited utility if considered in isolation.
Crucially, we carry out a set of experiments carefully designed to isolate the key irradiation damage variables and facilitate validation. Using tungsten as a model material, we find exceptionally good agreement between our numerical predictions and experimental measurements of defect densities and defect cluster sizes.

\end{abstract}


\end{frontmatter}
 
\section{Introduction}

The study of irradiation damage in crystalline materials and its accumulation and effects on material properties is an exceedingly complex problem that has attracted considerable attention over the last several decades \cite{weber2000models,woo2005modeling,wirth2007does,becquart2011modeling,dudarev2013density,xiao2020theoretical,fan2025atomistic}. 
Irradiation brings the system to a far-from-equilibrium state generally characterized by defect concentrations many orders of magnitude higher than in thermal conditions. This triggers a number of transformations that lead to \emph{microstructural evolution}, understood as a series of processes by which the system changes its internal structure to reduce its defect concentrations and lower its free energy.

A popular choice in the study of irradiation damage in metals is tungsten (W), which has attracted substantial attention over the last decade as a candidate plasma-facing material in fusion energy devices \cite{rieth2013recent,rieth2025tungsten,marian2017recent,YU2019297,qian2021using,mcelfresh2024fracture}. 
As such, considerable effort has been put on developing material models that can capture the fine-scale features of irradiation damage in W. 
Electronic structure calculations based on density functional theory (DFT) are typically employed to calculate fundamental material properties and defect energetics \cite{becquart2010microstructural,chen2012stability,boisse2014modelling,ventelon2012ab}, while classical molecular dynamics (MD) simulations are used to obtain transport properties and simulate collisional displacement events \cite{fikar2009molecular,nandipati2015displacement,liu2023large,liu2025utilizing}. Importantly, while atomistic simulations provide high physical accuracy, they do so at the expense of computational efficiency, eliminating them as a viable tool to reach irradiation doses of interest in experimental studies. Instead, simulations of high-dose irradiation in pure W are done using `object' kinetic Monte Carlo (okMC), or mean-field rate theory (MFRT) methods  \cite{becquart2010microstructural,MARIAN2012293,castin2017onset,marian2017recent,ZHANG2023101443,jin2018breaking,castin2018object,mason2019atomistic,hou2021influence,ma2024initial,wu2025influence,mohamed2025investigation}. These methods make use of computationally or experimentally obtained defect properties to surge above atomic-level time dynamics and extend the simulations' time reach near or above 1 dpa. Unfortunately, lacking clear experimental references to benchmark against, these studies are hampered by the need to incorporate the large number of irradiation mechanisms that have been proposed for W \cite{marian2017recent}. 
Fortunately, recent experimental studies have provided much needed information regarding the defect species present in the microstructure as a function of key irradiation variables \cite{schwendeman2025integral,hu2025new,zavavsnik2025microstructural, Schwendeman_2026}, thereby providing new opportunities to refine and improve mesoscale irradiation models.

Given the breadth and depth of irradiation effects in W that are captured in the models, it is of interest to carry out targeted experiments to isolate specific variables that may help us discriminate among the key mechanisms controlling the material response in each case. Accordingly, the main objective of this work is to perform a set of carefully designed irradiation experiments to help shed light on some of the most fundamental response mechanisms of irradiated W.
We intentionally use W specimens with minimal internal microstructure: specifically, well-annealed, very high-purity W single crystals with negligible intrinsic defect densities (such as dislocations) and low mosaicity (i.e., low-angle grain boundaries). This eliminates the need to consider empirical bias factors for defect absorption by sinks, which are traditionally a source of uncertainty in many studies of irradiation damage modeling, particularly in metals \cite{wolfer1975stress,malerba2007object,wu2025influence}. Furthermore, this enables X-ray diffuse scattering (XRDS) measurements \cite{schwendeman2025integral, Schwendeman_2026} to be used to quantify the irradiation-induced defect cluster densities and size distributions in the material, including clusters with sizes below the typical resolution limits of transmission electron microscopy (TEM) measurements, for experimental validation. Our simulations and experiments are also conducted at 290 K, below irradiation recovery Stage III in W, where vacancy motion is negligible. Collectively, these conditions present a unique opportunity to validate aspects of the physical model of irradiation damage in W pertaining specifically to the cascade-induced defect cluster distributions and the subsequent room-temperature thermal evolution of this defect population.

\section{Physical model of irradiation damage} \label{model}

\begin{figure}[ht]
\centering
\boxed{
  \includegraphics[width=0.97\linewidth, trim = 0 1.0cm 0 0cm, clip]{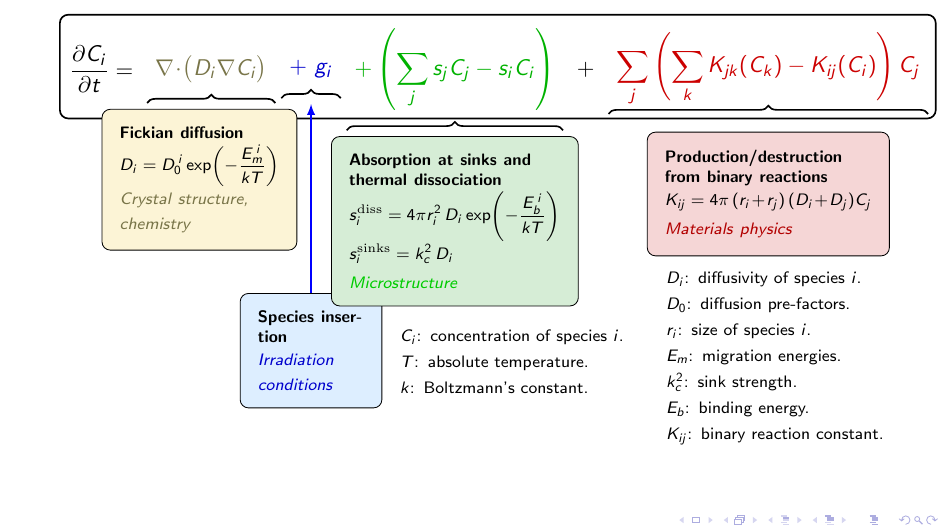} }
  \caption{Generalized mean-field cluster dynamics PDE based on classical nucleation theory. The right-hand side of the equation is broken down into separate contributions representing fundamentally different physical processes. These include diffusion, insertion of damage species, thermal evaporation of monomers from clusters, defect absorption at system sinks, and second-order cluster reaction kinetics. Each term is defined by specific material coefficients (diffusivities, dissolution probabilities, sink strengths, reaction constants, etc.). All these coefficients encode the physics of the process, the host material, and the microstructure. The expression for the reaction coefficient $K_{ij}$ in the red box labeled \emph{Materials Physics} represents the specific case of mutually-reacting 3D-moving objects. \label{fig:big-slide}}
\end{figure}

Mean-field rate theory provides the basis within which a system of partial differential equations (PDEs) representing the coupled cluster dynamics can be solved to simulate the evolution of irradiation damage in materials \cite{golubov2007kinetics,ghoniem2020rate,odette1988mechanisms}. Figure \ref{fig:big-slide} shows the generalized mean-field cluster dynamics PDE based on classical nucleation theory. As the figure illustrates, the equation is formulated as a linear superposition of process change rates, truncated to second-order defect interactions to simplify the many-body problem\footnote{This is generally a very good approximation, as the interaction cross sections for three-body reactions and above are extremely small \cite{korobeinikov1975calculation}.}. Numerical integration of this PDE system is feasible on modern computers, although to obtain practicable solutions, it is still necessary to apply certain numerical workarounds to accelerate the simulations \cite{xu2020grouping,cao2006efficient}.

Models as described in Fig. \ref{fig:big-slide} are useful because they are formulated in a general and flexible manner intended to accommodate our latest knowledge in irradiation effects coming from a number of more accurate simulation methods. These may include (i) correlated defect distributions from displacement cascades, (ii) irradiation phenomena such as radiation enhanced diffusion, (iii) radiation induced precipitation and segregation, (iv) multi-species chemistry, etc. This information is generally encoded in coefficient matrices with the same dimensionality as that of the the PDE system. However, the above system of equations is designed to capture the entire defect cluster population from $i=1,\ldots N$, where $N$ is the largest cluster size, which can in some cases reach millions of point defects. As such, the dimensionality of the cluster population can be extremely large, leading to large, potentially time-evolving, physical coefficient matrices that are sparse in some regions of the cluster dimensional space while concentrated in others, with elements that have widely disparate values. This complexity presents a significant numerical challenge that even the most robust deterministic solvers often struggle to overcome. 

The PDE system within the mean-field approximation can be solved by treating each term in the coefficient matrix as an event rate. In this fashion, one naturally introduces the stochastic variability of explicit stochastic methods such as the kinetic Monte Carlo (kMC) approach but, crucially, the mathematical form of MFRT methods is preserved. Event rates must be defined for a specific material volume, similar to kMC simulations, but matrix coefficients are obtained just as in MFRT simulations. Likewise, spatial derivatives (diffusion terms) can be converted to rates in the simulation volume using the divergence theorem with standard discretizations as in the deterministic case. We call this method \emph{stochastic cluster dynamics} (SCD) \cite{marian2011stochastic}, and a brief review of SCD is provided in \ref{app:scd}. For the remainder of this paper, the SCD method should be regarded as a stochastic version of the MFRT method: identical in its theoretical basis but more versatile and flexible in terms of the domain of application of the model and the numerical robustness of its solutions.

\subsection{Experimental}\label{sec:exp}

The experiments have been designed to facilitate comparison with the modeling predictions by considering the following points:   
\begin{enumerate}
\item Utilizing well-annealed, 99.999\% purity W single crystals precludes the presence of high-angle grain boundaries and minimizes the densities of both low-angle grain boundaries and intrinsic defects (e.g., dislocations and precipitates). These are typically the most potent defect sinks found in crystalline materials, and thus their absence simplifies the treatment of defect interactions. In addition, this eliminates the need to compute the coefficients $k_c^2$ shown in the third term in the r.h.s.~of Fig.\ \ref{fig:big-slide}, known as \emph{sink strengths}, which are often difficult to calculate and subject to a large degree of empiricism \cite{malerba2007object}.  
\item The irradiations were conducted at 290 K, which is below the Stage III temperature where vacancy mobility is effectively negligible. \cite{anand1978recovery,nambissan1992positron}. This leaves single self-interstitial atoms (SIA) and di-interstitials as the sole mobile species in the system \cite{dicarlo1969stage,KEYS1970260,anand1978recovery,seidman1979point,wilson1980situ}, which largely simplifies the defect kinetics in the simulations.
\item Irradiation with 10.8 MeV W self-ions ensures sufficient penetration to isolate the high-damage region of the specimen from its front and back surfaces, thereby eliminating spurious local effects resulting from undue proximity of defects to the free surfaces. Moreover, implanting with W self-ions eliminates potential chemical interactions between the ions and the host lattice.  
\end{enumerate}
The irradiation-induced defect cluster densities and size distributions were experimentally quantified using `integral' XRDS measurements \cite{schwendeman2025integral, Schwendeman_2026} around the surface-normal 110 Bragg reflection of the W single crystals. The analysis of these measurements is based on detailed modeling of the atomic structure and the lattice strain associated with irradiation-induced defects, as well as the diffuse scattering produced by statistical distributions of these defect clusters across macroscopic volumes of single crystals \cite{Dederichs_1971, Larson_1980, Ehrhart_1982, schwendeman2025integral, Schwendeman_dissertation, Schwendeman_2026}. Importantly, XRDS is distinguished by its ability to determine the densities of defect clusters with sizes below the typical resolution limits of TEM measurements \cite{Larson_1987, Olsen_2016}. In addition, due to the opposite signs of the long-range strain fields surrounding interstitial and vacancy-type defect clusters, XRDS can effectively extract separate size distributions for the two cluster types \cite{Larson_1987, Olsen_2016, Schwendeman_dissertation, Schwendeman_2026}. These two capabilities, in particular, provide a level of detail regarding the defect size distributions produced directly by the displacement cascades as well as the distinct defect clustering fractions, binding energies, and thermal mobilities of interstitial and vacancy clusters that enable direct comparison with SCD modeling predictions.

XRDS measurements around Bragg reflections are primarily sensitive to defect clusters that produce strong strain fields in the surrounding lattice \cite{Ehrhart_1994, Larson_2019, Schwendeman_dissertation}. In ion-irradiated W, TEM measurements \cite{YI2016105, Ferroni_2015, Guo_2020} and atomistic modeling \cite{byggmastar2025four} have demonstrated that these will be interstitial and vacancy-type dislocation loops with primarily $\left<111\right>$ Burgers vectors. The XRDS measurements have been analyzed accordingly \cite{Schwendeman_2026} to determine separate size distributions for these two loop types with loop diameters ranging from $0.6$ to 40 nm, which are comprised of $\sim$7 to over 10,000 point defects. Importantly, existing experimental studies \cite{Hollingsworth_2022, zavavsnik2025microstructural} and modeling \cite{Granberg_2021} have demonstrated that isolated vacancies (i.e. monovacancies) and very small globular (i.e. more spherically shaped) vacancy clusters are also produced under the irradiation conditions studied here (this is consistent with the experimental and modeling results presented in this work). These monovacancies and small vacancy clusters result in comparatively weak strain fields, and thus their densities and size distributions are not determined by the XRDS measurements. 


\subsection{Simulation conditions}\label{sec:srim}

\begin{figure}
    \centering
    \includegraphics[width=0.85\linewidth]{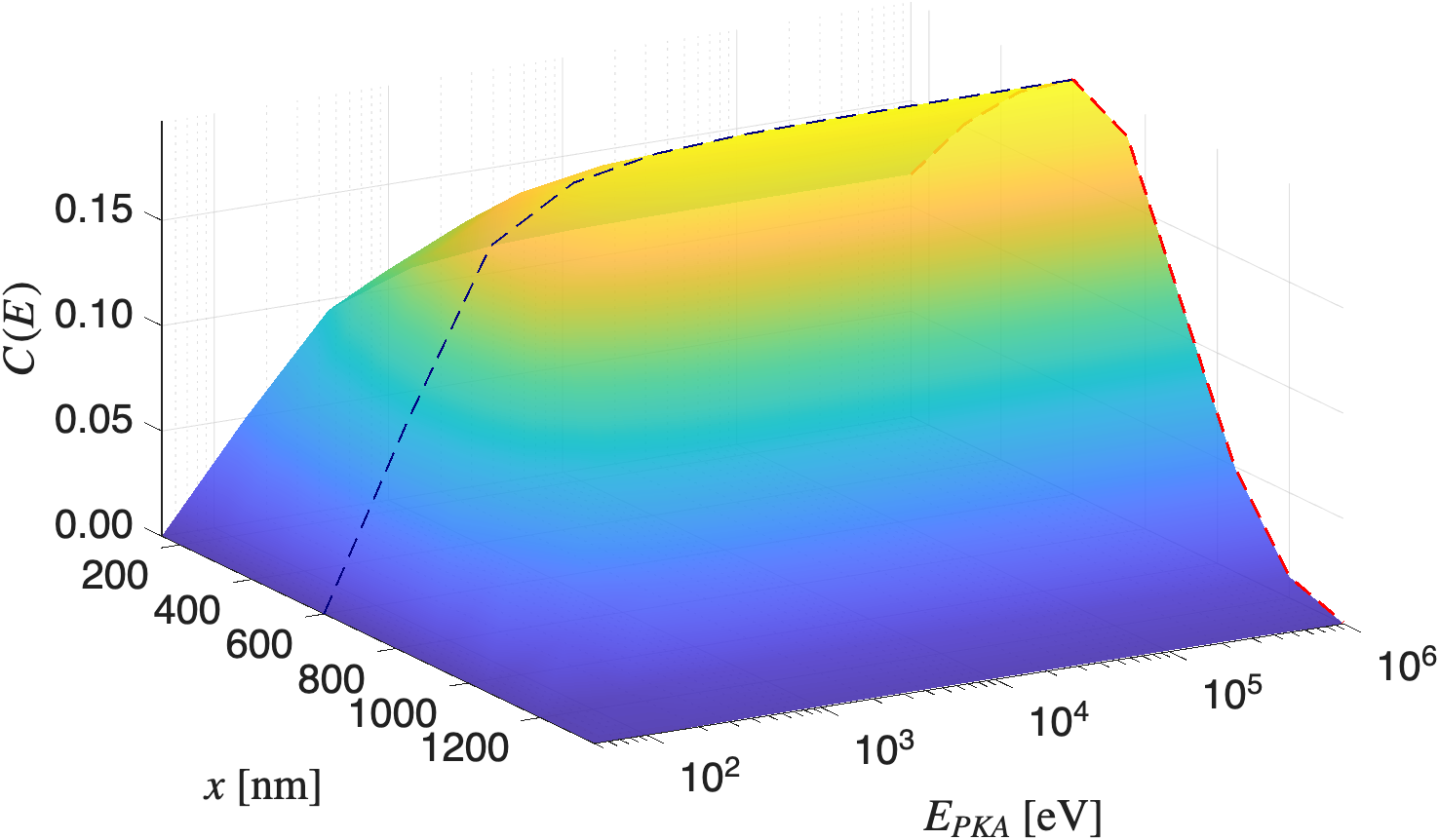}
    \caption{Cumulative PKA energy distribution function as a function of ion penetration depth due to 10.8 MeV W ion irradiation in W. The dashed black line represents the depth at which the maximum damage is attained ($x$=600 nm). The dashed red line indicates the depth damage profile at the point of saturation, which is seen to roughly follow a Bragg profile. The average PKA energy, obtained by integration of the two-dimensional surface in the entire $x$-$E$ space, is 0.8 keV.}
    \label{fig:cpdf}
\end{figure}

\begin{table}[H]
     \centering
     \caption{SCD simulation parameters defined from the experimental conditions considered in this work.}
     \begin{tabular}{|c|c|c|c|}
     \hline
          Parameter & Symbol & Value & Units \\
         \hline
         Ion fluence & - & $2.34\times10^{16}$ & m$^{-2}$ \\
         Irradiation time & $\Delta t_{\rm irr}$ & 93 & s \\
         Ion flux & - & $2.52\times10^{14}$ &  m$^{-2}$$\cdot$s$^{-1}$ \\
         Total dose & - & 0.008 & dpa \\
         Dose rate & $\dot\gamma$ & $8.6\times10^{-5}$ & dpa$\cdot$s$^{-1}$ \\
         Temperature & $T$ & 290 & K \\
         Atomic density & $\rho_a$ & $6.3\times10^{28}$ & m$^{-3}$\\
         Burgers vector & $b$ & 0.27 & nm \\
         Numerical volume & $\Omega$ & $10^{-18}$ & m$^3$\\
         Annealing time & - & 3600 & s \\
         \hline
     \end{tabular}
     \label{tab:param}
\end{table}
The starting point for simulating irradiation damage processes in solids is the PKA (primary knock-on atom) energy distribution function that describes the energies with which recoils are produced during particle collisions. For 10.8 MeV self-ion irradiations of pure W used in the experiments performed by Schwendeman and co-workers \cite{schwendeman2025integral, Schwendeman_2026}, the PKA distribution has been extracted from \texttt{SRIM} simulations in full cascade mode \cite{SRIM}. 
Figure \ref{fig:cpdf} shows the cumulative PKA energy distribution function, $C(x;E)$, as a function of depth beneath the sample's surface. The \texttt{SRIM} calculations indicate that the average energy expended on lattice damage out of the total 10.8 MeV of ion energy is 2.65 MeV. Additionally, the average number of displacements per ion track is $N_{\rm NRT}$=18,373. The mean PKA energy is then obtained from Fig.\ \ref{fig:cpdf} as:
\begin{equation}
   \langle E\rangle = \int_0^L\int_0^{\infty}C(x;E)~dx~dE
\end{equation}
which reflects the depth dependence of recoil production in ion irradiation experiments through the variable $x$, extending from the front surface at $x=0$ to the back surface at $x=L$, as well as its dependence on recoil energy $E$. For the present case, we obtain a value of $\langle E\rangle=0.8$ keV, although it is important to note that $\langle E\rangle$ is not a representative substitute of the full PKA energy spectrum, as we will demonstrate below. Damage accumulates at a rate of $8.6\times10^{-5}$ dpa$\cdot$s$^{-1}$ \footnote{We use the common unit of \emph{displacements per atom} or `dpa' to represent the damage fluence. In this study with 10.8 MeV self-ion irradiation of W, 1 dpa amounts to a dose of approximately $2.9\times10^{18}$ ions per m$^2$ \cite{schwendeman2025integral}.}, and the rest of the simulation parameters are given in Table \ref{tab:param}. 

In the configuration used here, XRDS measurements are not depth-sensitive, and thus the XRDS results provide a depth-integrated picture of the defect size distributions. The relative defect density is assumed to follow the SRIM-calculated damage-depth profile illustrated in Fig.\ \ref{fig:cpdf}, and the defect densities reported by the experimental analysis correspond to the peak damage depth at $x=600$ nm \cite{schwendeman2025integral, Schwendeman_2026}. Therefore, for cross-consistency between experiments and modeling, here we sample PKA energies from the entire $C(x;E)$ function, i.e., first we sample the depth variable, $x^*$, with a probability given by the $x$ dependence of $C(x)$, followed by an energy sampling of $C(x^*;E)$.
In order to match the experimental conditions, the simulations also include a one-hour annealing period at 290 K. This annealing is simulated by allowing the defects present in the system after irradiation to evolve according to their thermal relaxation laws (see terms with Boltzmann factors in Fig.\ \ref{fig:big-slide} and eq.\ \eqref{eq:scd}). Thermal annealing is generally controlled by diffusion of mobile species and dissociation of large immobile species by emission of monomers as described in the following subsection and the Appendix.

\subsection{Key physics ingredients of the model}


The model captures several crucial pieces of irradiation defect physics that have proved essential to reproduce experimental measurements. Some of these are based on newly-acquired understanding from very recent works:
\begin{enumerate}
    \item Using cascade-induced defect distributions obtained directly from MD simulations, which do not account for post-cascade correlated recombination.
    \item Using a universal fractal law for the sizes of defect clusters directly emerging from displacement cascades, regardless of whether they are of SIA or vacancy nature. 
    \item Assuming that the only mobile species are single self-interstitials, di-interstitials, and monovacancies.
    \item Capturing the fast one-dimensional character of SIA diffusion at low temperature.
    \item Considering sufficiently large simulation volumes to appropriately sample the recoil energy distribution's high energy tail.
\end{enumerate}
The following subsections discuss each of these items in detail as they have been implemented in the model.

\subsubsection{Defect production in high-energy displacement cascades}
The quantity $\tilde{g}$ in eq.\ \eqref{eq:scd} represents the insertion rate of defect species in the system. Defects are introduced in discrete bursts, each representing an independent displacement cascade, which are generated using correlations extracted from extensive databases produced by molecular dynamics (MD) simulations of high-energy displacement cascades in W \cite{setyawan2015displacement,ZHANG2023101443,marian2025computational,byggmastar2025four}. The relevant information provided by these databases includes (i) the total number of Frenkel pairs, $N_{\rm PKA}$, produced by a PKA with a given energy, (ii) the fraction of defects that form part of clusters and (iii) the distribution of the size of the clusters. 
Here we use the following expressions: 
\begin{equation}
N_{\rm PKA}=\Bigg\{\begin{matrix}
    3.81E^{0.62}, & E\leq48~{\rm keV} \\
    0.50E^{1.15}, & E>48~{\rm keV} 
\end{matrix}
\label{eq:corr}
\end{equation}
These correlations are all dynamically sampled by the SCD code in every iteration and defects are introduced according to the PKA energy distribution provided in Fig.\ \ref{fig:cpdf}. Note that some of the aforementioned MD studies are very recent (and supersede previous studies \cite{troev2011simulation,sand2013high}) and thus only available to the nuclear materials community in the last few months.

\subsubsection{Cluster size distribution}
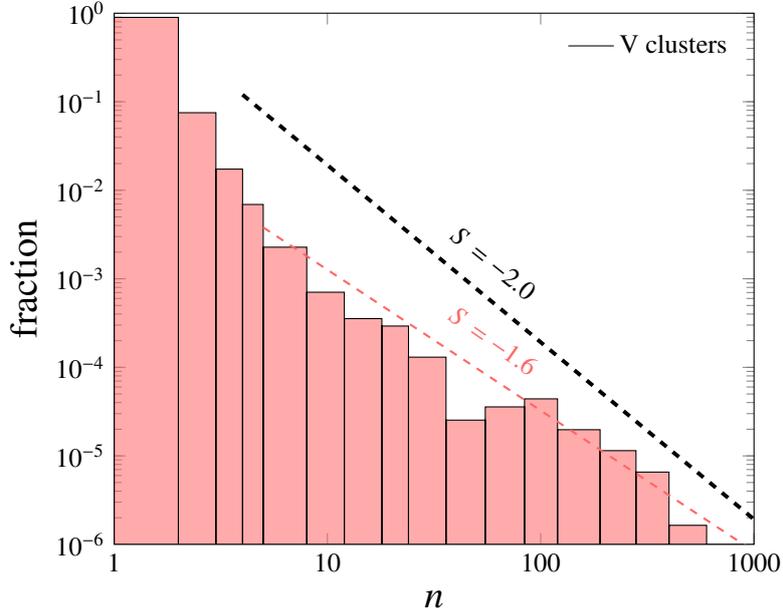
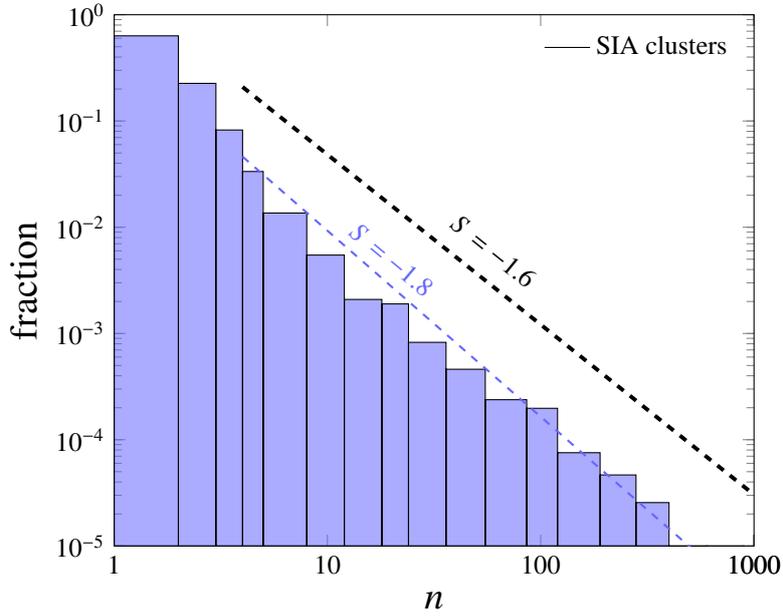
\begin{figure}[htp!]
\centering
\subfigure[Vacancy clusters\label{vac-dist}]{
\begin{tikzpicture}
\begin{axis}[
  xmode=log, log basis x=10,
  ymode=log, log basis y=10,
  xmin=1, xmax=1001,
  ymin=0.000001, ymax=1,
  log origin=infty,
  enlarge x limits=false,        
  enlarge y limits=false,
  xlabel={\Large $n$},
  ylabel={\Large fraction},
  xtick={1.0001,10,100,1000},
  xticklabels={1,10,100,1000},
  tick align=inside,
  legend style={at={(0.98,0.98)},anchor=north east,draw=none,fill=none},
]


\addplot+[
  ybar interval, mark=none,
  draw=black, fill=red!60, fill opacity=0.55
] table[
  x=Size,
  y=V-clusters,
  col sep=comma
]{figures_Cluster-dist_V-dist.csv};
\addlegendentry{V clusters}



\addplot+[no marks, domain=5:1000, samples=150, color=red!60, thick, dashed, fill=none]
  {0.0492 * x^(-1.59)};
\node[rotate=-35,color=red!60]  at (axis cs:60,0.0002) {$S=-1.6$};

\addplot+[no marks, domain=4:1000, samples=150, color=black, line width=1.5pt, dashed, fill=none]
  {1.923 * x^(-2.0)};
\node[rotate=-39,color=black]  at (axis cs:60,0.0015) {\bf $S=-2.0$};

\end{axis}
\end{tikzpicture}
}
\subfigure[SIA clusters\label{sia-dist}]{
\begin{tikzpicture}
\begin{axis}[
  xmode=log, log basis x=10,
  ymode=log, log basis y=10,
  xmin=1, xmax=1001,
  ymin=0.00001, ymax=1,
  log origin=infty,
  enlarge x limits=false,        
  enlarge y limits=false,
  xlabel={\Large $n$},
  ylabel={\Large fraction},
  xtick={1.0001,10,100,1000,1000},
  xticklabels={1,10,100,1000,1000},
  tick align=inside,
  legend style={at={(0.98,0.98)},anchor=north east,draw=none,fill=none},
]


\addplot+[
  ybar interval, mark=none,
  draw=black, fill=blue!60, fill opacity=0.55
] table[
  x=Size,
  y=SIA-clusters,
  col sep=comma
]{figures_Cluster-dist_SIA-dist.csv};
\addlegendentry{SIA clusters}



\addplot+[no marks, domain=4:1000, samples=150, color=blue!60, thick, dashed, fill=none]
  {0.523 * x^(-1.751)};
\node[rotate=-40,color=blue!60]  at (axis cs:20,0.005) {$S=-1.8$};

\addplot+[no marks, domain=4:1000, samples=150, color=black, line width=1.5pt, dashed, fill=none]
  {1.923 * x^(-1.6)};
\node[rotate=-39,color=black]  at (axis cs:60,0.006) {\bf $S=-1.6$};

\end{axis}
\end{tikzpicture}
}
\caption{Cluster size distributions for W from high-energy displacement cascade simulations by Byggmastar \etal~\cite{byggmastar2025four}. \protect\subref{vac-dist} Vacancy clusters. \protect\subref{sia-dist} SIA clusters. 
Note that the fraction in each cluster size bin has been normalized by the bin width such that the area under the histograms is equal to one. Inverse power law exponents are shown next to dashed segments, indicating the numerical scaling expected for cluster sizes with $n>4$. The correlations extracted from the work of Sand \etal~\cite{sand2013high} are shown as black dashed lines for comparison.}
\label{fig:barchart}
\end{figure}

While the total number of Frenkel pairs per PKA is the starting point of the defect insertion process, another key piece of information concerns the sizes of the defect clusters inserted in the simulation volume by each displacement cascade. 
Detailed studies based on MD simulations have suggested that cluster sizes follow a universal scaling law independent of PKA energy \cite{sand2013high,mason2019atomistic}. This is supported by recent results by Byggmastar \etal~\cite{byggmastar2025four} including cascades with PKA energies up to 2 MeV.  
The normalized histograms based on the full dataset are shown in Figure \ref{fig:barchart} for vacancy, Fig.\ \ref{vac-dist}, and SIA, Fig.\ \ref{sia-dist}, clusters. We have 
fitted the distributions to inverse power laws with scaling $\sim$$n^{S}$, where $n$ is the number of point defects in the cluster and $S$ is the scaling exponent.
As the plots show, the exponents are $-1.6$ and $-1.8$ for vacancy and SIA clusters, denoted as $V_n$ and $I_n$, respectively. We use these values in our simulations to allocate the number Frenkel pairs generated by each PKA using the damage correlations in eqs.\ \eqref{eq:corr} to defect clusters, until the total number of defects per cascade is fully apportioned. 

\subsubsection{Mobile defect species}\label{sec:mob}

MD simulations going back to the 1990s have established that the general outcome of a dense displacement cascade is a `shell' of  SIA clusters surrounding a vacancy-rich core. SIA clusters above a certain size appear as dislocation loops, which in body-centered cubic (BCC) metals are crystallographically `perfect' (unfaulted) and thus potentially glissile along their Burgers vector's direction \cite{wirth2000dislocation,marian2003md}. However, in W, SIA clusters are practically immobile for sizes larger than three interstitials \cite{faney2014spatially}. For their part, vacancies can also appear forming dislocation loops, although their most likely configuration below $n$<$30$ is as globular clusters or nanovoids \cite{yang2022combined,hu2025new,de2023readdressing,becquart2011modeling,hu2021effect}. Similar to their SIA counterparts, vacancy loops are effectively immobile, and, except for select exceptions \cite{wei2025resolving}, atomistic calculations suggest that only monovacancies possess sufficient mobility among all vacancy clusters \cite{oda2014first,wu2024high,long2016first}.

In fact, for W there is consistent evidence from isochronal annealing studies that the only species contributing to recovery in stages I, II, and III are, respectively, single SIA, di-interstitials, and monovacancies \cite{dicarlo1969stage,KEYS1970260,dausinger1975long,anand1978recovery,seidman1979point,wilson1980situ,nambissan1992positron}. TEM experiments in high-purity electron irradiated W have shown that clusters move only in highly driven conditions, which are distinct from conventional ion beam irradiation conditions  \cite{arakawa2020quantum}. Even in atomistic simulations, in pure lattices with no impurities or imperfections, it has been shown that small SIA clusters in W have a reduced mobility compared to other BCC metals \cite{zhou2014creeping}. Object kinetic Monte Carlo simulations also point to limited or no SIA cluster mobility to find consistency with experimental defect concentration measurements \cite{jin2018breaking}.

Despite some contradicting work on the diffusion properties of single self-interstitials, in which migration energies as high as 0.3 eV have been reported \cite{chen2012stability,zhou2013dynamical,long2017study,wang2021migration}, we have found that the migration energies of SIAs that lead to agreement with the present experiments (and are consistent with additional experimental measurements \cite{amino2016detection}) should be $<$0.01 eV. One of the main reasons for these low SIA migration energies is that their migration is almost exclusively one-dimensional along $\langle111\rangle$ directions \cite{amino2016detection}. It is known that considering rotation as part of the transport mechanism of SIAs leads to larger effective migration energies, as several atomistic calculations have proposed \cite{derlet2007multiscale,chen2012stability,long2017study,zhou2024self}. However, this is at odds with experimental observations \cite{amino2011activation,amino2016detection,heikinheimo2019direct} (and other conflicting atomistic studies \cite{derlet2007multiscale,becquart2010microstructural,faney2014spatially}). Moreover, some authors have suggested that the reason for this low migration energy is that quantum effects may suppress the temperature dependence at low temperatures in W \cite{swinburne2017low}. 
With that in mind, the diffusivity parameters for the mobile species in our model, and the sources for their values, are listed in Table \ref{tab:mob}.
\begin{table}[H]
   \centering
\caption{Diffusivity parameters for the mobile defects employed in this work. \label{tab:mob}}
    \begin{tabular}{|c|c|c|c|}
    \hline
        Defect & $D_0$ [m$^2$$\cdot$s$^{-1}$] & $E_m$ [eV] & Source \\
    \hline
        $I_1$ & $8.77\times10^{-8}$ & 0.009 & \cite{faney2014spatially,heikinheimo2019direct} \\
        $I_2$ & $7.97\times10^{-8}$ & 0.024 & \cite{faney2014spatially} \\
        $V_1$ & $1.11\times10^{-6}$ & 1.71 & \cite{mundy1978self,ventelon2012ab,hossain2014stress,heikinheimo2019direct,maksimenko2022n} \\
    \hline
    \end{tabular}
 \end{table}

\subsubsection{One-dimensional migration of single self-interstitials}

As indicated in the previous subsection, SIA mobility is a key property governing the agreement between the experimental measurements and the modeling predictions. With the migration properties firmly established, here we characterize its other crucial aspect: 1D diffusion\footnote{Note that in the context of irradiation damage, `1D diffusion' refers to the migration of the defects along rectilinear paths in a 3D setting.}. Consideration of one-dimensional migration of self-interstitials implies employing material parameters in the equation shown in Fig.\ \ref{fig:big-slide} and in eq.\ \eqref{eq:scd} that reflect this characteristic of SIA motion. In particular, 1D migration leads to specific coefficients $\tilde{s}_i$ and $\tilde{K}_{ij}$. In our setting, where the only operational defect sinks are free surfaces bounding the irradiated sample film, the corresponding sink strength, $k_c^2$, for SIA migration is \cite{borodin1998rate,huang2019rates}:
$$k^2_c=\frac{4}{l^2}$$
where $l$ is the thickness of the irradiated film (approximately 1 mm for the experiments here). The reaction rate between a 1D-moving species $i$ (e.g., an SIA) and an immobile object $j$ (a vacancy, vacancy cluster, or another SIA cluster) is written as \cite{huang2019rates}:
\begin{equation}
K_{ij}=8 \pi D_i\left(r_i+r_j\right)^{2} C_iC_j^{4/3}
    \label{rij}
\end{equation}
where $D_i$ is the diffusivity of (mobile) species $i$; $r_i$ and $r_j$ are the interaction radii of the two reacting species; and $C_i$ and $C_j$ are their respective concentrations. Note that the above rate closely resembles that of the 3D-3D case for low values of $C_j$. Following Huang and Marian \cite{huang2019rates}, the reaction rate between two one-dimensionally migrating species reactions (e.g., single SIAs reacting with one another) is neglected in this work due to their extremely low interaction cross-sections.

\subsubsection{PKA energy distribution tail}

The final element in our model that has a strong bearing on the ability of our simulations to capture the experimental size distributions is the need to introduce PKAs with sufficiently high energies to sample the large-size portion of the cluster size distribution. As justified above, due to the limited defect mobility at 290 K in W, the only feasible formation mechanism of large loops at this temperature is through direct injection by displacement cascades. 
However, as noted by Sand \etal~\cite{sand2013high} and Byggmastar \etal~\cite{byggmastar2025four}, the probability of generating loops with more than 100 defects in W requires PKA energies higher than 100 keV (which is consistent with the value obtained using eq.\ \eqref{eq:corr}). From Fig.\ \ref{fig:cpdf}, the probability of sampling energies larger than 100 keV is approximately one in 100,000. Because each ion produces on average 18,373 recoils, roughly $100,000/18,373\approx5.5$ ion insertions are needed to produce one 100 keV cascade. Thus, to ensure that the high-energy part of the PKA energy distribution is adequately sampled, a sufficient number of ion insertion events must be simulated ($\gtrsim6$). The number of such events can be estimated from the ion insertion rate, $\dot G$, in the simulation volume defined as:
\begin{equation}
    \dot{G}=\frac{\dot{\gamma}\rho_a\Omega}{N_{\rm NRT}}
    \label{eq:dotg}
\end{equation}
where $\dot{\gamma}$ is the dpa rate, $\rho_a$ the atomic density of W, and $\Omega$ the simulation volume described in \ref{app:scd} (all these parameters are given in Table \ref{tab:param}). For sufficient statistics, it is desirable to maintain a high ion insertion rate, which, in the context of our simulations, implies using as large a value of $\Omega$ as reasonably possible. This imposes numerical demands on the simulations that have to be confronted with appropriate computational resources, but that are essential to adequately sample the large-size defect concentration tails observed experimentally.
In this work, we use $\Omega=10^{-18}$ m$^{-3}$, i.e., one cubic micron.


\section{Results}
\subsection{Cluster size distributions}\label{sec:clustersize}

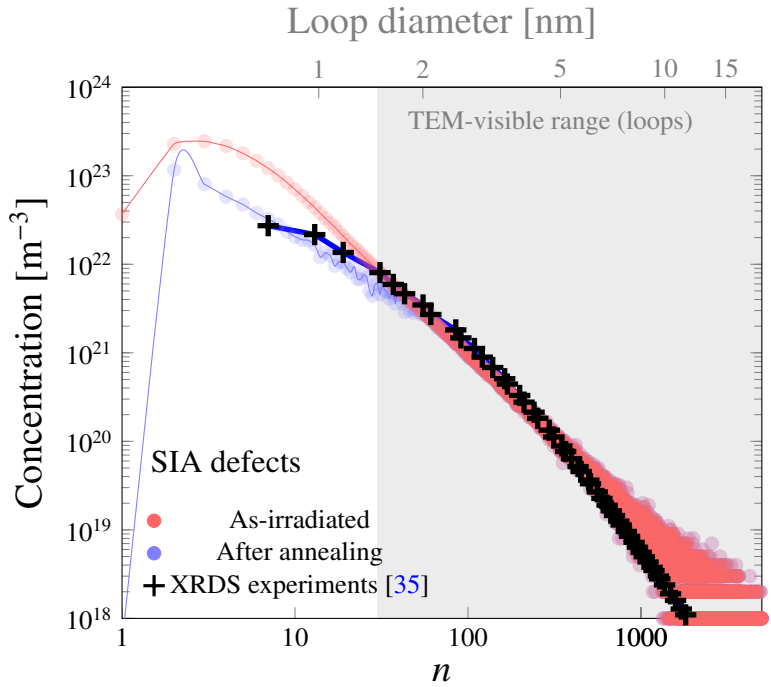
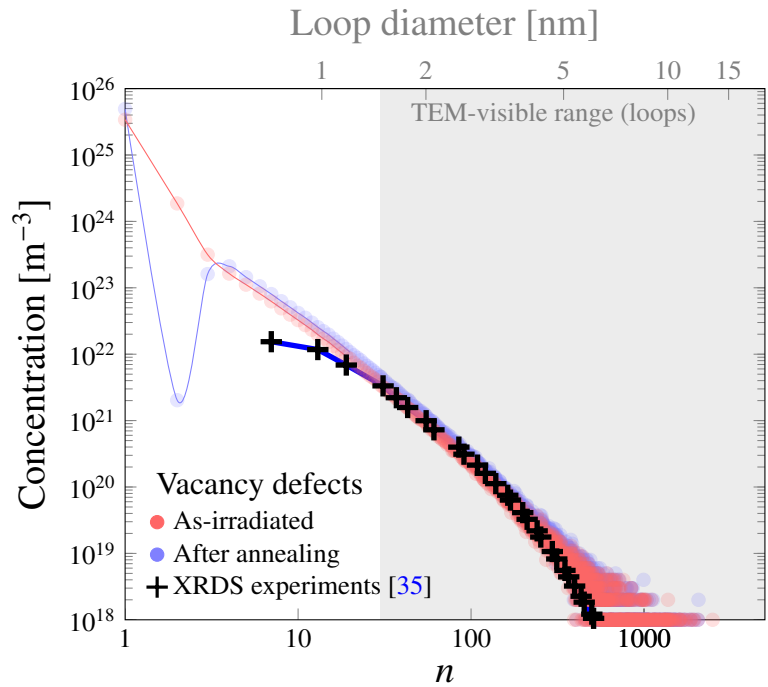
\begin{figure}[htp]
\centering
\subfigure[SIA clusters\label{sia-scatter}]{


 




\begin{tikzpicture}
\begin{axis}[
name=base,
  set layers,
  xmode=log, log basis x=10,
  ymode=log, log basis y=10,
  xmin=1, xmax=5000,
  ymin=1e18, ymax=1e24,
  log origin=infty,
  xtick pos=bottom,
  enlarge x limits=false,        
  enlarge y limits=false,
  xlabel={\Large $n$},
  ylabel={\Large Concentration [m$^{-3}$]},
  xtick={1.0001,10,100,1000,1000},
  xticklabels={1,10,100,1000,1000},
  every mark/.append style={solid,draw=none},
  tick align=inside,
  legend style={at={(0.02,0.02)},anchor=south west,draw=none,fill=none},
]

\begin{pgfonlayer}{axis background}
  \fill[gray!15] (axis cs:30,1e18) rectangle (axis cs:5000,1e24);
\end{pgfonlayer}
\node[color=gray] at (axis cs:300,4e23) {TEM-visible range (loops)};

\addplot+[only marks,
    mark=*,
    mark size=2.5pt,
    mark options={draw=blue!50!white, fill=blue!50!white, opacity=0.2},
    opacity=0.4,
    forget plot,] 
table[
  x=size,
  y expr=\thisrow{concentration}*1e+6,
  col sep=comma
]{figures_Cluster-scatter_sia_after_anneal.csv};
\addplot+[smooth,                
  tension=0.4, 
  no markers,           
  color=blue!50!white,  
  forget plot
] table[
    col sep=comma,
    x expr=(\thisrow{size}>40) ? nan : \thisrow{size},
    y expr=\thisrow{concentration}*1e+6,
  ] {figures_Cluster-scatter_sia_after_anneal.csv};

\addplot+[only marks,
  mark=*,
  mark size=2.5pt,
  mark options={fill=red!60, draw=red!60},  
  opacity=0.2,
  forget plot,
] table[x=size, y expr=\thisrow{concentration}*1e+6, col sep=comma]
      {figures_Cluster-scatter_sia_before_anneal.csv};
\addplot+[smooth,                
  tension=0.2,
  no markers,           
  color=red!60,  
  forget plot
] table[
    col sep=comma,
    x expr=(\thisrow{size}>40) ? nan : \thisrow{size},
    y expr=\thisrow{concentration}*1e+6,
  ] {figures_Cluster-scatter_sia_before_anneal.csv};

\addplot+[line width=2pt,
  mark=+,
  mark size=4pt,
  mark options={fill=black, draw=black, line width=2pt},  
forget plot,
] table[
  x=size,
  y expr=\thisrow{concentration}*1e+6,
  col sep=comma
]{figures_Cluster-scatter_sia_exp.csv};

\addlegendimage{only marks, mark=*, mark size=2.5pt,
  mark options={fill=red!60, draw=red!60}}
\addlegendentry{As-irradiated}

\addlegendimage{only marks, mark=*, mark size=2.5pt,
  mark options={draw=blue!50!white, fill=blue!50!white}}
\addlegendentry{After annealing}

\addlegendimage{only marks, mark=+, very thick, mark size=4pt, draw=black}
\addlegendentry{XRDS experiments \protect\cite{Schwendeman_2026}}




\node[font=\large, color=black] at (axis cs:4,6e19) {SIA defects};

\end{axis}

\begin{axis}[
        at={(base.south west)}, anchor=south west,   
        axis x line*=top,
        axis y line=none,
        xmode=log, log basis x=10, 
        xmin=0.27, xmax=19.1, 
        xlabel={\Large Loop diameter [nm]}, 
        xlabel style={color=gray},
        error bars/y dir=both, error bars/y explicit,
        yticklabel=\empty,
        ymode=normal, ymin=0, ymax=1, 
        xticklabel style={color=gray},
        scaled x ticks=false,
        xtick={1,2,5,10,15},  
        x tick scale label code/.code={}, 
        xticklabel style={/pgf/number format/fixed,/pgf/number format/precision=2,/pgf/number format/fixed zerofill},
        xticklabels={1,2,5,10,15}
        ]
      \end{axis}

\end{tikzpicture}

}
\subfigure[Vacancy clusters\label{vac-scatter}]{
\begin{tikzpicture}
\begin{axis}[
name=base,
  set layers,
  xmode=log, log basis x=10,
  ymode=log, log basis y=10,
  xmin=1, xmax=5000,
  ymin=1e18, ymax=1e26,
  log origin=infty,
  enlarge x limits=false,        
  enlarge y limits=false,
  xlabel={\Large $n$},
  ylabel={\Large Concentration [m$^{-3}$]},
  xtick={1.0001,10,100,1000,1000},
  xticklabels={1,10,100,1000,1000},
  xtick pos=bottom,
  every mark/.append style={solid,draw=none},
  tick align=inside,
  legend cell align=left,
  legend style={at={(0.02,0.02)},anchor=south west,draw=none,fill=none},
]

\begin{pgfonlayer}{axis background}
  \fill[gray!15] (axis cs:30,1e18) rectangle (axis cs:5000,1e26);
\end{pgfonlayer}
\node[color=gray] at (axis cs:300,4e25) {TEM-visible range (loops)};

\addplot+[only marks,
    mark=*,
    mark size=2.5pt,
    mark options={draw=blue!50!white, fill=blue!50!white, opacity=0.2},
    opacity=0.4,
    forget plot] 
table[x=size, y expr=\thisrow{concentration}*1e+6, col sep=comma]
      {figures_Cluster-scatter_vac_after_anneal.csv};
\addplot+[smooth,            
  no markers,           
  color=blue!50!white,  
  forget plot
] table[
    col sep=comma,
    x expr=(\thisrow{size}>20) ? nan : \thisrow{size},
    y expr=\thisrow{concentration}*1e+6,
  ] {figures_Cluster-scatter_vac_after_anneal.csv};

\addplot+[only marks,
  mark=*,
  mark size=2.5pt,
  mark options={fill=red!60, draw=red!60},  
  opacity=0.2,
  forget plot
] table[
  x=size,
  y expr=\thisrow{concentration}*1e+6,
  col sep=comma
]{figures_Cluster-scatter_vac_before_anneal.csv};

\addplot+[smooth,            
  no markers,           
  color=red!60,  
  forget plot
] table[
    col sep=comma,
    x expr=(\thisrow{size}>20) ? nan : \thisrow{size},
    y expr=\thisrow{concentration}*1e+6,
  ] {figures_Cluster-scatter_vac_before_anneal.csv};

\addplot+[line width=2pt,
  mark=+,
  mark size=4pt,
  mark options={fill=black, draw=black, line width=2pt},  
forget plot,
] table[
  x=size,
  y expr=\thisrow{concentration}*1e+6,
  col sep=comma
]{figures_Cluster-scatter_vac_exp.csv};




\addlegendimage{only marks, mark=*, mark size=2.5pt,
  mark options={fill=red!60, draw=red!60}}
\addlegendentry{As-irradiated}

\addlegendimage{only marks, mark=*, mark size=2.5pt,
  mark options={draw=blue!50!white, fill=blue!50!white}}
\addlegendentry{After annealing}

\addlegendimage{only marks, mark=+, very thick, mark size=4pt, draw=black}
\addlegendentry{XRDS experiments \protect\cite{Schwendeman_2026}}



\node[font=\large, color=black] at (axis cs:6,1e20) {Vacancy defects};

 \end{axis}

\begin{axis}[
        at={(base.south west)}, anchor=south west,   
        axis x line*=top,
        axis y line=none,
        xmode=log, log basis x=10, 
        xmin=0.27, xmax=19.1, 
        xlabel={\Large Loop diameter [nm]}, 
        xlabel style={color=gray},
        error bars/y dir=both, error bars/y explicit,
        yticklabel=\empty,
        ymode=normal, ymin=0, ymax=1, 
        xticklabel style={color=gray},
        scaled x ticks=false,
        xtick={1,2,5,10,15},  
        x tick scale label code/.code={}, 
        xticklabel style={/pgf/number format/fixed,/pgf/number format/precision=2,/pgf/number format/fixed zerofill},
        xticklabels={1,2,5,10,15}
        ]
      \end{axis}

\end{tikzpicture}
}
\caption{Simulated defect cluster size distributions at end of irradiation (0.008 dpa) and after post-irradiation annealing for one hour at room temperature. \protect\subref{sia-scatter} SIA clusters. \protect\subref{vac-scatter} Vacancy clusters. The dislocation loop size distributions determined from the XRDS measurements are plotted for comparison in both cases. The shaded region represents the range of visible clusters by conventional TEM. For loops, this limit ($n=30$) is obtained by setting the diameter of a circular disc to a minimum diameter of 1.5 nm.}
\label{fig:scatter}
\end{figure}

Figure \ref{fig:scatter} shows the computed SIA and vacancy cluster size distributions at the end of the irradiation period and after post-irradiation annealing for one hour at room temperature. In both cases, the dislocation loop size distributions determined experimentally by the XRDS measurements (corresponding to the `after annealing' simulation results) are shown for comparison. For reference, the figure also shows the extent of TEM-visible cluster size range assuming planar loops, i.e., $n>30$.

Figure \ref{sia-scatter} displays the SIA cluster size distributions in the `as irradiated' and `after annealed' conditions. The scatter in the model defect size distributions is due to the stochastic nature of our simulations, which, for a given parameter set, yield different but statistically valid results when repeated several times. Furthermore, the statistical variance increases with cluster size due to limited statistics. 
As the figure shows, the difference between the as-irradiated and annealed size distributions is most visible for cluster sizes with $n<30$. This is a direct consequence of the high mobility of $I_1$ and $I_2$ defects, whose populations, in the absence of irradiation, are partitioned in the following manner:
\begin{enumerate}
    \item[(i)] A portion is consumed via recombination with vacancies and vacancy clusters.
    \item[(ii)] Another fraction is absorbed by existing SIA loops, which contributes to their steady --albeit modest-- growth during the annealing. 
    \item[(iii)] Emission of single SIA from small SIA clusters, enabled by modest binding energies of clusters with $n<10$ (see Table \ref{tab:bind}).
\end{enumerate}
Our results show that the modeled defect size distributions match the experimental results, both before and after post-irradiation annealing, for larger cluster sizes ranging from $n = 30$ to as large as $n \sim 1000$, where the statistical variance in the modeled distribution becomes significant. Interestingly, the agreement is remarkably accurate if one considers the lower envelope (convex hull) of the simulated statistical distributions. For smaller defects, including the smallest loop size ($n = 7$) incorporated into the experimental analysis, the agreement improves after the simulated annealing, owing principally to a decline in the populations of small SIA clusters through dissociation and recombination.

Figure \ref{vac-scatter} shows the vacancy cluster size distributions for the as-irradiated and annealed conditions. 
Qualitatively, the vacancy cluster population displays a size distribution characteristically similar to its self-interstitial counterpart for larger cluster sizes ($n > 30$). However, consistent with the experimental loop size distributions \cite{Schwendeman_2026}, the larger vacancy cluster densities predicted by the model are substantially lower than the SIA densities of the same size, and the vacancy size distribution falls off significantly faster with increasing $n$. Thus, the majority of the defects in the TEM-visible size range will be SIA clusters. Another important distinction from the SIA distribution is that the plot shows a stable monovacancy contribution in the as-irradiated condition that reaches very high concentrations. This is a consequence of the negligible mobility of vacancies at 290 K, which leads to a high retention of monovacancies in the material. After the annealing, most of the di-vacancy ($V_2$) population disappears on account of its thermal instability (negative binding energy, see Table \ref{tab:bind}), and a concentration maximum develops for the $V_{4}$ cluster.
For smaller vacancy cluster sizes ($n < 30$), the model predicts appreciably higher densities with decreasing cluster size than the experiment. This can be attributed to the fact that the XRDS measurements are analyzed to extract the vacancy loop size distribution, specifically (as discussed in Sec.~\ref{sec:exp}), while the modeled cluster distribution includes both vacancy loops and globular vacancy clusters.

\subsection{Defect cluster evolution}\label{sec:evolution}

\begin{figure}[H]
\centering
\subfigure[SIA-type defects\label{sia-concentration}]{
\resizebox{0.516\textwidth}{!}{


\begin{tikzpicture}

\begin{axis}[
  name=base,
  xmode=log, log basis x=10,
  ymode=log, log basis y=10,
  xmin=0.04, xmax=118,
  ymin=5e19, ymax=1e26,
  xtick pos=bottom,
  xtick={0.1,1,10,100},      
  tick align=inside,
  xlabel={\Large time [s]},
  ylabel={\Large Concentration [m$^{-3}$]},
    legend style={
    at={(0.14,0.98)},
    anchor=north west,
    draw=none,
    fill=none,
    font=\large
  },
]

  \addplot+[no marks, line width=2pt, color=blue!40!white]
  table[x=time, y=concentration, col sep=comma]
  {figures_concentration_evolution_sia1.csv};
\addlegendentry{$n=1$}

\addplot+[no marks, line width=2pt, color=red!70]
table[x=time, y=concentration, col sep=comma]
{figures_concentration_evolution_sia2.csv};
\addlegendentry{$n>2$}

\addplot+[no marks, line width=3pt, dashed, color=green!80!black]
table[x=time, y=concentration, col sep=comma]
{figures_concentration_evolution_sia30.csv};
\addlegendentry{$n>30$}



\node[font=\large, color=black] at (axis cs:0.07,1e25) {\Large $I_n$};

\end{axis}


\begin{axis}[
        at={(base.south west)}, anchor=south west,   
        axis x line*=top,
        axis y line=none,
        xmode=log, log basis x=10, 
        xmin=0.00000344, xmax=0.010148, 
        xlabel={\Large dose [dpa]}, 
        error bars/y dir=both, error bars/y explicit,
        yticklabel=\empty,
        ymode=normal, ymin=0, ymax=1, 
        scaled x ticks=false,
        xtick={0.00001,0.0001,0.001,0.01},  
        x tick scale label code/.code={}, 
        xticklabel style={/pgf/number format/fixed,/pgf/number format/precision=2,/pgf/number format/fixed zerofill},
        xticklabels={$10^{-5}$,$10^{-4}$,$10^{-3}$,$10^{-2}$},
           legend style={at={(0.1,0.02)},
                    anchor=south west,
                    draw=none,
                    fill=none,
                    font=\large
  },
        ]

      \end{axis}

\end{tikzpicture}




}
}
\subfigure[Vacancy-type defects\label{vac-concentration}]{
\resizebox{0.452\textwidth}{!}{


\begin{tikzpicture}

\begin{axis}[
  name=base,
  xmode=log, log basis x=10,
  ymode=log, log basis y=10,
  xmin=0.04, xmax=118,
  ymin=5e19, ymax=1e26,
  xtick pos=bottom,
  xtick={0.1,1,10,100},      
  tick align=inside,
  xlabel={\Large time [s]},
    legend style={
    at={(0.14,0.98)},
    anchor=north west,
    draw=none,
    fill=none,
    font=\large
  },
  yticklabel=\empty
]

  \addplot+[no marks, line width=2pt, color=blue!70!white]
  table[x=time, y=concentration, col sep=comma]
  {figures_concentration_evolution_vac1.csv};
\addlegendentry{$n=1$}

\addplot+[no marks, line width=2pt, color=red!70]
table[x=time, y=concentration, col sep=comma]
{figures_concentration_evolution_vac2.csv};
\addlegendentry{$n>2$}

\addplot+[no marks, line width=3pt, dashed, color=green!80!black]
table[x=time, y=concentration, col sep=comma]
{figures_concentration_evolution_vac33.csv};
\addlegendentry{$n>30$}


\node[font=\large, color=black] at (axis cs:0.07,1e25) {\Large $V_n$};


\end{axis}


\begin{axis}[
        at={(base.south west)}, anchor=south west,   
        axis x line*=top,
        axis y line=none,
        xmode=log, log basis x=10, 
        xmin=0.00000344, xmax=0.010148, 
        xlabel={\Large dose [dpa]}, 
        error bars/y dir=both, error bars/y explicit,
        yticklabel=\empty,
        ymode=normal, ymin=0, ymax=1, 
        scaled x ticks=false,
        xtick={0.00001,0.0001,0.001,0.01},  
        x tick scale label code/.code={}, 
        xticklabel style={/pgf/number format/fixed,/pgf/number format/precision=2,/pgf/number format/fixed zerofill},
        xticklabels={$10^{-5}$,$10^{-4}$,$10^{-3}$,$10^{-2}$}
        ]
      \end{axis}

\end{tikzpicture}




}
}
\caption{Evolution with time of the concentrations of \protect\subref{sia-concentration} SIA ($I_n$) and \protect\subref{vac-concentration} vacancy ($V_n$) clusters. Each plot tracks the evolution of single SIAs and monovacancies (blue curves), clusters with more than two defects (red), and clusters with sizes above the TEM resolution limit (dashed green): defined here as circular dislocation loops with $n>30$. 
}
\label{fig:concentration}
\end{figure}
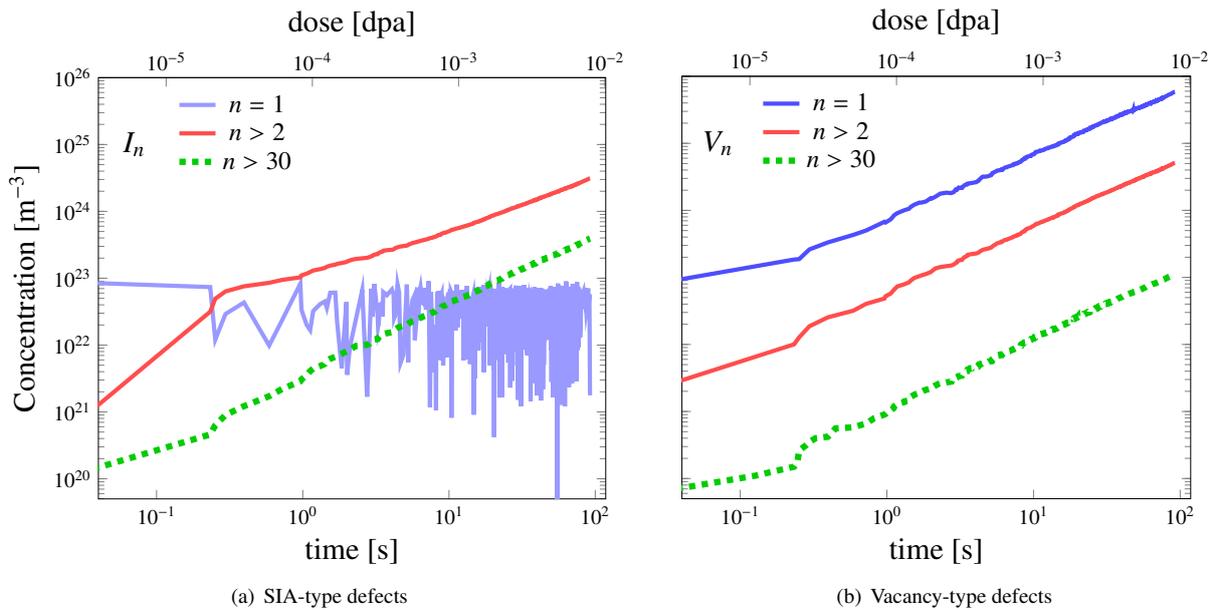

Figure \ref{fig:concentration} shows the evolution of the SIA and vacancy subpopulations as a function of irradiation time. Each population is partitioned into single point defects ($I_1$ and $V_1$), all clusters ($n>2$), and prismatic loops larger than the TEM visibility limit ($n>30$). The evolution of SIA defects is displayed in Fig.\ \ref{sia-concentration}, where the unstable nature of single interstitials is clear from its relatively low concentrations and highly volatile time evolution. This is, as discussed above, a direct consequence of their high mobility and transient behavior. In contrast, the SIA cluster populations exhibit well-defined, monotonically-increasing trends throughout the entire irradiation process. 

The evolution of the vacancy clusters is shown in Fig.\ \ref{vac-concentration}. In this case, the monovacancy population is seen to be stable and grow at the same rate as larger clusters, although in much higher concentrations. Indeed, more than 90\% of vacancies appear and remain as monovacancies in the simulations. The concentration of visible loops is 500 times smaller than that of monovacancies. 
Clusters of both types are observed to increase linearly, indicating that the system is found in a transient stage of defect accumulation and has not reached saturation yet. A comparison of these results with existing time-resolved experimental data is provided in Sec.\ \ref{sec:disc-heavy}.

It is interesting to note that the fraction of large SIA clusters ($n>30$) increases at a faster rate than the general SIA cluster population ($n>2)$. This is not the case for vacancy clusters, clearly showcasing the increasing importance of SIA loops in acting as effective sinks for single SIAs and small SIA clusters as they progressively accumulate in the material with irradiation dose. This also highlights the time-dependent nature of the defect size distributions, which must be taken into account when interpreting defect cluster concentration evolutions.

\section{Discussion}

\subsection{Capabilities of physics-based numerical simulation model}

We have identified four essential factors without which model predictions would not have achieved the high degree of agreement with the experimental data. In no particular order, these factors are: (i) accurately capturing the amount of defects produced in cascades as a function of PKA energy; (ii) a universal scaling law of cluster sizes for defects generated in displacement cascades (independent of PKA energy); (iii) one-dimensional, athermal SIA diffusion;  and (iv) consideration of the disproportionate importance of the high-energy tail of the PKA energy distribution function. All four have been implemented in our models, jointly contributing to the accuracy of the simulations and the agreement with experiments. Point (i) ensures that a sufficient number of defects are inserted to account for the large clusters observed experimentally. Point (ii) guarantees that the defect clusters inserted into the simulation volume appear with the correct sizes based on the primary damage mechanisms unveiled by MD studies. Point (iii) maintains a high SI mobility at low temperatures, which controls recombination with vacancies and post-cascade SIA loop growth. Finally, point (iv) plays the crucial role of ensuring that the material is seeded with few but important large immobile dislocation loops, which act as a stable substrate on which further irradiation damage accumulates and evolves. Another factor related to this last point is the importance of considering the full PKA energy distribution, not just the average PKA energy (or any other single representative energy) as has been done in several studies \cite{zhang2017radiation,Granberg_2021}.

The factors above pertain to the physics of irradiation damage in W, and are in principle agnostic to the model through which they are implemented. These may include okMC \cite{becquart2010microstructural,becquart2011modeling,wu2025influence}, MFRT \cite{Li_Zhou_Ning_Huang_Zeng_Ju_2012,jin2018breaking,li2022towards}, or our hybrid method SCD, which builds on advantageous features of both: (i) a mean-field description of the defect populations for computational efficiency and reaching high irradiation doses, and (ii) uses a kinetic Monte Carlo solver to avoid numerical solution bottlenecks in MFRT and provide a meaningful statistical variance to assess numerical errors.

Key to the model validity is the realization that most of the energetics employed are obtained with atomistic simulations. Both semi-empirical potential and DFT-based atomistic calculations have reached an impressive degree of accuracy, which carries over to the models and decisively contributes to their validity.

\subsection{Comparison between model predictions and XRDS measurements}

The simulated SIA and vacancy defect cluster distributions show excellent overall agreement with the XRDS measurements across two orders of magnitude of cluster sizes spanning nearly four orders of magnitude in concentration. For clusters with sizes $n > 30$, the modeling predictions match the experimental results both before and after post-irradiation annealing at 290 K. The distinct SIA and vacancy cluster size distributions measured in this larger size range can therefore be primarily attributed to the differing formation mechanisms of the two cluster types within the cascades. As shown by \cite{byggmastar2025four}, both clusters types with $n > 30$ are produced in cascades initiated by $> 50$ keV PKAs. While Fig. \ref{fig:barchart} shows that the inverse power law size distributions for the two cluster types are very similar, the clustering fraction for vacancies remains substantially lower than that for SIAs in these high energy cascades \cite{byggmastar2025four}. This crucial distinction is further illustrated by Fig. \ref{fig:concentration} and is ultimately reflected in the differing cluster distribution predictions at the end of the irradiation period. The remarkable agreement between these modeling predictions and the XRDS-measured distributions shown in Fig. \ref{fig:scatter} represents an important experimental validation of the general accuracy of recent cascade physics studies.

For SIA defects (Fig.\ \ref{sia-scatter}) in the as-irradiated condition, the concentrations of clusters with $n<30$ diverge from the measured distribution with increasingly higher predicted cluster concentrations with decreasing cluster size. This discrepancy is mitigated after post-irradiation annealing at 290 K, suggesting that the combined effect of thermal coarsening and monomer dissociation during a prolonged annealing period substantially affects the concentrations of smaller cluster sizes present in the material over experimental time scales. This has important implications for comparing model predictions with experimental measurements of defect distributions and property evolution---measurements that are often performed ex situ at room temperature, as was the case in this study. The evolution of the vacancy cluster distribution during post-irradiation annealing is more subtle, with Fig.\ \ref{vac-scatter} showing that the only substantial change in the modeling predictions is the reduction of the concentration of the di-vacancy population by several orders of magnitude due to its thermal instability as discussed above. Instead, the significantly lower clustering fraction for vacancies within the cascades (for all PKAs with energies $> 1$ keV \cite{byggmastar2025four}) results in a stable and extremely high concentration of monovacancies. 

Fig.\ \ref{vac-scatter} shows that the model also predicts increasingly higher concentrations of vacancy clusters with decreasing size in comparison to the experimental distribution for $n < 30$. This discrepancy is attributed to the analysis of the XRDS measurements extracting the concentrations of vacancy loops, specifically. Importantly, these XRDS measurements do not preclude the presence of small globular vacancy clusters, which have weak distortion fields and therefore do not produce strong XRDS intensities around Bragg reflections. In fact, as discussed by Schwendeman \etal~\cite{Schwendeman_2026}, 85\% of the point defects in the XRDS-measured dislocation loop population are found to be SIAs. Assuming roughly equal densities of SIA and vacancy point defects under these irradiation conditions, the large density of `missing' vacancies in the loop population is consistent with PAS measurements (see the following subsection) and the modeling results shown here that indicate a large population of (primarily) monovacancies and small globular clusters is present in the material. Furthermore, by taking the difference between the number of point defects in the XRDS-measured interstitial and vacancy loop populations, a concentration of approximately $4\times10^{25}$ m$^{-3}$ vacancy point defects is inferred to be in non-loop structures. This agrees remarkably well with the modeling prediction of approximately $5\times10^{25}$ m$^{-3}$ monovancacies after post-irradiation annealing, as illustrated in Fig.\ \ref{vac-scatter}. 

The experimental interstitial and vacancy loop size distributions shown in Fig. \ref{fig:scatter} have been determined from least-squares fits to XRDS measurements using size-dependent cross-section calculations for dislocation loops \cite{Schwendeman_dissertation, Schwendeman_2026}. Such XRDS cross-section calculations are fundamentally based on detailed atomistic modeling of the structures of defect clusters and the distortions they induce in the surrounding lattice \cite{Ehrhart_1982, Schwendeman_dissertation}. Like most studies to date, the cluster modeling utilized in the analysis here involves manually constructing the dislocation loops at the center of a lattice and then imposing elastic continuum theory calculations of the defect-induced displacement field on the surrounding atoms \cite{Schwendeman_dissertation}. However, it is well known that the structure and distortion fields of the smallest defect clusters ($n \lesssim 20$) can diverge appreciably from this continuum theory based modeling \cite{marian2002100}. As discussed in detail by Larson \cite{Larson_2019} and further by Schwendeman and coworkers \cite{Schwendeman_dissertation, Schwendeman_2026}, this motivates incorporating MD modeling of the defect clusters into the XRDS cross-section calculations, which can can, in principle, be extended down to point defects and the very smallest cluster sizes with arbitrary configurations. This would enable improved and more detailed comparisons between the modeling and measurements in the smallest cluster size range. Nevertheless, the agreement observed in Fig. \ref{fig:scatter} with the existing experimental analysis, particularly for larger cluster sizes ($n > 20$), is outstanding, which adds a high level of confidence to the underlying physical principles behind the simulation models. 
Future work will investigate scenarios involving different irradiation conditions --specifically temperature, particle energy, and particle mass-- with the ultimate goal of extrapolating these findings to fusion energy conditions in leading conceptual designs.

\subsection{Discussion on literature on heavy ion irradiation of W}\label{sec:disc-heavy}

A large body of heavy-ion irradiation studies in W have been performed at temperatures near the operational window for fusion applications, $>$500$^\circ$C. Under such conditions, monovacancies are mobile, leading to the formation of vacancy clusters and voids of various sizes depending on the irradiation conditions.
Irradiations with 18 MeV self-ions at 500 and 800$^\circ$C up top 0.2 dpa by Hwang \etal~\cite{HWANG2016430} revealed both dislocation loops and voids, with average sizes of, respectively, 3.1 and 1.0$\sim$1.5 nm. 
These vacancy clusters are at or below the visibility threshold in TEM, which hampers our ability to extract statistically meaningful counts of these defects. This limitation can be addressed using positron annihilation spectroscopy (PAS) (e.g., refs. \cite{debelle2008first,hu2025new}), which can detect down to monovacancies. Consistent with the aforementioned TEM results, recent PAS measurements by Zava\v{s}nik \etal~\cite{zavavsnik2025microstructural} of W crystals irradiated with 10.8 MeV self-ions to 0.02 and 0.2 dpa at 800 K similarly report the presence of larger vacancy clusters comprised of 25$\sim$50 point defects (approximately 1 nm in diameter) at elevated irradiation temperatures. Interestingly, despite the presence of larger clusters, new detailed PAS analyses of W irradiated with 18-MeV self-ions to 0.0085 dpa nevertheless indicate that approximately 98\% of the total concentration of vacancy clusters consists of defects with $n<5$ \cite{hu2025new}. Although these irradiations were performed at, or around, 500$^\circ$C, the presence of large $>3$ nm dislocation loops and large densities of small vacancy clusters are consistent with our results in Fig.\ \ref{fig:scatter}.

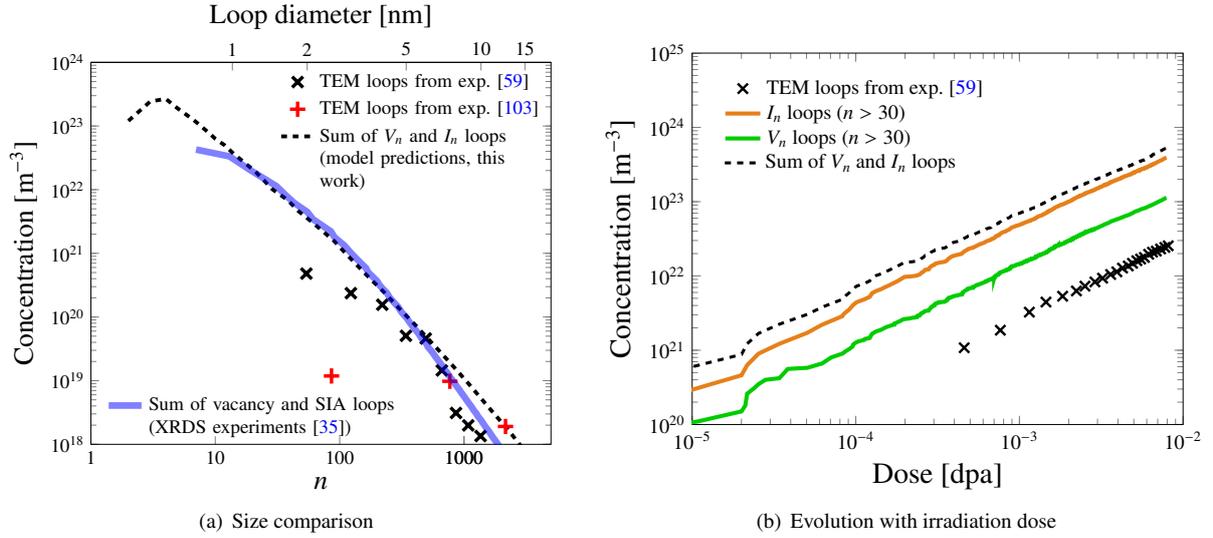
\begin{figure}[htb]
\centering
\subfigure[Size comparison\label{compa-size}]{
\resizebox{0.46\textwidth}{!}{





\begin{tikzpicture}
\begin{axis}[
name=base,
  set layers,
  xmode=log, log basis x=10,
  ymode=log, log basis y=10,
  xmin=1, xmax=5000,
  ymin=1e18, ymax=1e24,
  log origin=infty,
  xtick pos=bottom,
  enlarge x limits=false,        
  enlarge y limits=false,
  xlabel={\Large $n$},
  ylabel={\Large Concentration [m$^{-3}$]},
  xtick={1.0001,10,100,1000,1000},
  xticklabels={1,10,100,1000,1000},
  every mark/.append style={solid,draw=none},
  tick align=inside,
  legend style={
  at={(1.03,0.99)},
  anchor=north east,
  draw=none,
  fill=none,
  text width=4.3cm, 
  nodes={inner sep=2pt, anchor=west},
  row sep=2pt
  },
]


\pgfplotstableread[col sep=comma]{
Size,Concentration
54, 480662804229712.4
123, 236828521833592.53
220, 155828325176946.5
341, 50688266332685.78
492, 45983532672501.21
663, 14493007739961.334
862, 3110066089982.9375
1082, 1984242101285.2454
1359, 1349748036306.8477
}\datatable
\addplot[
    only marks,
    mark=x,
    mark size=4pt,
    mark options={line width=1.8pt},
    color=black
]
table[
    col sep=comma,
    y expr =\thisrow{Concentration}/1.0e-6,
    x= Size,
]{\datatable};
\addlegendentry{TEM loops from exp.~\cite{YI2016105}} 

\pgfplotstableread[col sep=comma]{
Size,Concentration
86, 11845643224479295000
770, 9773714352524533000
2156, 1906856850800013000
4242, 129220675789042070
}\datatabless
\addplot[
    only marks,
    mark=+,
    mark size=4pt,
    mark options={line width=1.8pt},
    color=red, 
]
table[
    col sep=comma,
    y expr =\thisrow{Concentration},
    x= Size,
]{\datatabless};
\addlegendentry{TEM loops from exp.~\cite{Wang_2023}}

\pgfplotstableread[col sep=comma]{
Size,Concentration
2, 120351815327545890
3, 243209588978134560
4, 267505883547608420
5, 187880378430055070
6, 145315286767391330
7, 116046280231890560
8, 92683848637846850
9, 76444511539130500
10, 63050503732210640
11.809821840052303, 50354111665400264
14.069833995200819, 36526272639656344
21.047180053023766, 19216169478539748
28.845365456651297, 10112547403434790
38.172363380817444, 6877636884957037
47.09814498574782, 4531137520189450
60.18214747926508, 3082043968903186
90.02696793763015, 1570304202847029.2
115.03672306357413, 939613777070405.1
163.27803180905016, 463721476068172.4
227.72696001100422, 244019482864261.38
296.13032941314634, 141399256905821.62
413.01857290378734, 72060607478505.88
586.2202772156909, 34442070515497.39
803.4206051255165, 17553612242914.412
1120.545917832982, 8390431390194.262
1562.846591140141, 3884056862110.791
2297.294802616453, 1632894000906.1914
3040.1130514405772, 806067954539.6061
3884.6653529730347, 385408893474.73883
4963.8367551622205, 196473874647.40884
5611.118358991583, 117613071929.57645
}\datatables
\addplot[no marks, line width=2.0pt, color=black, dashed]
table[
    col sep=comma,
    y expr =\thisrow{Concentration}/1.0e-6,
    x= Size,
]{\datatables};
\addlegendentry{Sum of $V_n$ and $I_n$ loops (model predictions, this work)}

\end{axis}

\begin{axis}[
    at={(base.south west)}, anchor=south west,
    xmode=log, log basis x=10,
  ymode=log, log basis y=10,
  xmin=1, xmax=5000,
  ymin=1e18, ymax=1e24,
  log origin=infty,
    hide axis,              
    stack plots=y,          
    legend style={
  at={(0.018,0.06)},
  anchor=south west,
  draw=none,
  fill=none,
  text width=4.5cm, 
  nodes={inner sep=2pt, anchor=west}
  },
]
\addplot[draw=none] 
table[
  x=size,
  y expr=\thisrow{concentration}*1e+6,
  col sep=comma, forget plot
]{figures_Cluster-scatter_vac_exp.csv};

    \addplot[no marks, line width=3.5pt, color=blue, opacity=0.5]
    table[
  x=size,
  y expr=\thisrow{concentration}*1e+6,
  col sep=comma
]{figures_Cluster-scatter_sia_exp.csv};
\addlegendentry{Sum of vacancy and SIA loops (XRDS experiments \cite{Schwendeman_2026})}

\end{axis}

\begin{axis}[
        at={(base.south west)}, anchor=south west,   
        axis x line*=top,
        axis y line=none,
        xmode=log, log basis x=10, 
        xmin=0.27, xmax=19.1, 
        xlabel={\Large Loop diameter [nm]}, 
        xlabel style={color=black},
        error bars/y dir=both, error bars/y explicit,
        yticklabel=\empty,
        ymode=normal, ymin=0, ymax=1, 
        xticklabel style={color=black},
        scaled x ticks=false,
        xtick={1,2,5,10,15},  
        x tick scale label code/.code={}, 
        xticklabel style={/pgf/number format/fixed,/pgf/number format/precision=2,/pgf/number format/fixed zerofill},
        xticklabels={1,2,5,10,15}
        ]
      \end{axis}

\end{tikzpicture}

}
}
\subfigure[Evolution with irradiation dose\label{compa-dose}]{
\resizebox{0.50\textwidth}{!}{

\begin{tikzpicture}

\begin{axis}[
    name=base,
    width=10cm, height=8cm,
    xmode=log, ymode=log,
    xmin=1e-5, xmax=0.01,
    ymin=1e20, ymax=1e25,
    xlabel={\Large Dose [dpa]},
    ylabel={\Large Concentration [m$^{-3}$]},
    legend style={at={(0.05,0.95)}, anchor=north west, draw=none, fill=none},
    legend cell align={left}
]

\addplot+[only marks,
    mark=x,
    mark size=3.5pt,
    mark options={draw=black, very thick, fill=white},
    ] table[
            x expr=\thisrow{fluence}*0.01*55/90.0e+16,
            y expr=\thisrow{a_dens}/50.0e-9,
            col sep=comma
            ]
{figures_concentration_evolution_yi_tem_concentration.csv};
\addlegendentry{TEM loops from exp.~\cite{YI2016105}} 

    \addplot[no marks, line width=2pt, color=orange!80!gray]
    table[x expr=\thisrow{time}*8.6e-5, y=concentration, col sep=comma] {figures_concentration_evolution_sia30.csv};
    \addlegendentry{$I_n$ loops ($n>30$)}

    \addplot[no marks, line width=2pt, color=green!80!black]
    table[x expr=\thisrow{time}*8.6e-5, y=concentration, col sep=comma] {figures_concentration_evolution_vac33.csv};
    \addlegendentry{$V_n$ loops ($n>30$)}


\end{axis}

\begin{axis}[
    at={(base.south west)}, anchor=south west,
    width=10cm, height=8cm,
    xmode=log, ymode=log,
    xmin=1e-5, xmax=0.01,
    ymin=5e19, ymax=1e25,
    hide axis,              
    stack plots=y,          
    legend style={at={(0.05,0.75)}, anchor=north west, draw=none, fill=none}
]
    \addplot[forget plot, draw=none] 
    table[x=time, y=concentration, col sep=comma] {figures_concentration_evolution_vac60.csv};

    \addplot[no marks, line width=1.5pt, color=black, dashed]
    table[x expr=\thisrow{time}*8.6e-5, y=concentration, col sep=comma] {figures_concentration_evolution_sia30.csv};
    \addlegendentry{Sum of $V_n$ and $I_n$ loops}
\end{axis}



\end{tikzpicture}
}
}
\caption{\protect\subref{compa-size} Comparison between model predictions and XRDS \cite{Schwendeman_2026} results (after 10.8-MeV self-ion irradiation to 0.008 dpa at 290 K) with TEM-measured loop size distributions by  Yi \etal~\cite{YI2016105} (150-keV W ion irradiation to 0.01 dpa at 300 K) and Wang \etal~\cite{Wang_2023} (6.0-MeV Cu ion irradiation to 0.05 dpa at room temperature). \protect\subref{compa-dose} Evolution with dose of `TEM-visible' loop concentrations from simulations performed in this work and the in-situ TEM measurements by Yi \etal~\cite{YI2016105}. Note that Yi \etal~used a threshold displacement energy of 55 eV to calculate their dose rate, instead of the recommended value of 90 eV. For that reason, we have converted their dose points to an energy of 90 eV to enable a direct comparison with our results.}
\label{fig:comparison}
\end{figure}


Recent studies of ion-irradiated W employing a combination of TEM and PAS \cite{Wang_2024} have demonstrated that the larger TEM-visible voids observed in the studies referenced above at elevated temperatures result from the thermal evolution of the defect population due to mobilized monovacancies. PAS studies of ion-irradiated W near room temperature, below the Stage III recovery temperature \cite{debelle2008first, Keys_1970, Wang_2024}, have thus reported the presence of large densities of monovacancies and very small vacancy clusters, with no voids visible in TEM imaging \cite{Wang_2023, Wang_2024}. Hollingsworth \etal~\cite{Hollingsworth_2022} performed PAS measurements after 2-MeV self-ion irradiation to 0.0085 dpa at room temperature and reported a population of primarily monovacancies with an estimated concentration of $9.2 \times 10^{25}$ m$^{-3}$. This result agrees remarkably well with the modeling predictions and XRDS-based inference of the monovacancy density (see Fig.~\ref{vac-scatter} in the previous subsection) for the higher-energy ions used here. With increasing irradiation dose, this same work noted a trend toward increasing contributions from small $n < 7$ clusters to the measured spectra. Zava\v{s}nik \etal~\cite{zavavsnik2025microstructural} also performed PAS measurements of W single crystals irradiated at room temperature with 10.8 MeV self-ions, nearly identical to the conditions studied in this work. At both 0.02 and 0.2 dpa, they reported a population of primarily monovacancies and very small, $2 < n < 4$ vacancy clusters, in general agreement with the previous work and the modeling results presented here in Fig. \ref{fig:scatter}.

TEM imaging for room temperature irradiations of W using 6.0 MeV Cu ions by Wang and co-workers~\cite{wang2023defect, Wang_2023, Wang_2024} over a dose range of 0.05 to 0.6 dpa revealed a microstructure dominated by dislocation loops with an average size of 6$\sim$8 nm ($n\approx490\sim870$) at concentrations on the order of $10^{22}$ m$^{-3}$. Despite the higher irradiation dose range in comparison to that studied here, this loop concentration is over an order of magnitude lower than the TEM-visible SIA cluster concentration in Fig.\ \ref{sia-scatter} (assumed to be dislocation loops and defined in the figure as $n>30$) at the end of the irradiation period, corresponding to 0.008 dpa. Nevertheless, Figure \ref{fig:comparison} illustrates that the absolute dislocation loop size distribution measured by Wang \etal~\cite{Wang_2023} at the most comparable dose of 0.05 dpa is remarkably consistent with the results of this work for larger loop sizes $> 6$ nm. No loop nature distinction was made in their work, so the results are compared directly in Fig.~\ref{compa-size} to the total cluster size distribution obtained by adding the SIA cluster and vacancy cluster concentrations in Figs.\ \ref{sia-scatter} and \ref{vac-scatter} for both the SCD modeling predictions and the XRDS measurements. The observed discrepancy for smaller loop sizes $<$ 6 nm is consistent with a detailed study comparing XRDS and TEM measurements of defect cluster size distributions in irradiated nickel by Olsen \etal~\cite{Olsen_2016}, which illustrates that the TEM resolution limit is not sharp. Instead, for clusters $< 6$ nm in diameter, the TEM measurements obtained using the bright field imaging method utilized by Wang \etal~\cite{Wang_2023} were shown to increasingly undercount the defect cluster concentrations with decreasing size. Note that 6-nm loops in W correspond to $n \approx 490$, which is much larger than the smallest loop sizes that are often visible in TEM measurements that are undercounted due to resolution effects. Importantly, Wang \etal~\cite{Wang_2023} also pointed out that the loop microstructure saturates after an irradiation dose of 0.2 dpa, which is not seen here due to our lower doses, as demonstrated by the monotonically increasing trends in Fig.\ \ref{fig:concentration}. Finally, we note that their finding that the main vacancy defects were monovacancies and small clusters after irradiation at room temperature is consistent with the studies discussed above and the present calculations. 

A separate body of work by Yi \etal~\cite{YI2016105} involving 150-keV self-ion irradiations of W thin foils at room temperature reports the accumulation of TEM-visible dislocation loops in the amounts shown in Figure \ref{fig:comparison}. 
These authors reported a predominance of vacancy loops at low doses ($\lesssim$ 0.01 dpa) and room temperature, with a tendency towards equipartition as the dose increases ($>$ 0.1 dpa). These observations are consistent with an accompanying study done using 2-MeV W-ion irradiations in semi-bulk tungsten specimens \cite{yi2015characterisation}. However, establishing the defect nature of loops at low doses is based mostly on qualitative analysis of `black-and-white' contrast of TEM micrographs, and on the erroneous assumption that SIA loops are formed by diffusion and agglomeration of individual SIA defects, and can therefore only appear at high doses after sufficient SIA production has taken place \cite{haussermann1972study}. For that reason, here we make no assumptions about their nature and plot their concentrations directly in Figure \ref{compa-size} next to the total cluster size distributions for both the SCD modeling predictions and the XRDS measurements as described above. Despite the differing irradiation conditions, the figure shows remarkable agreement between the TEM-measured loop size distributions by Yi \etal~\cite{YI2016105} and those obtained in this work for loops greater than 4 nm in diameter. The improved agreement at smaller loop sizes in comparison to the Wang \etal~\cite{Wang_2023} TEM results is consistent with the higher resolution weak beam dark field imaging method used in the study by Yi \etal~\cite{YI2016105}, which Olsen \etal~\cite{Olsen_2016} demonstrated undercounts defect clusters smaller than 3$\sim$4 nm in diameter. 

Likewise, Fig.~\ref{compa-dose} shows the buildup of visible loops (i.e., with diameters larger than 1.5$\sim$2.0 nm) with irradiation dose as predicted by the SCD simulations (SIA and vacancy loops combined) and those measured experimentally using in-situ TEM in the study by Yi \etal~\cite{YI2016105}. 
In the range of doses common to our study and theirs, there is approximately a factor of $\times$20 difference in total concentrations. This is not surprising given the differences in the irradiation species, W specimens  (i.e. thin foils versus bulk), and the unsharp nature of the TEM resolution limit discussed above. Nonetheless, our simulations provide an almost exact match of the time exponents ($\sim$$t^k$, $k$$\approx$1.1) of the cluster population observed experimentally.


Finally, we discuss the nature and Burgers vectors of loops formed directly in displacement cascades in irradiated W. Prismatic loops may be of either self-interstitial or vacancy character, and can carry Burgers vectors of $\nicefrac{1}{2}\langle111\rangle$ or $\langle100\rangle$. This gives rise to four distinct loop types that may form under irradiation and must therefore be distinguished. The distinction is important because the energetics and mobilities of each type can be drastically different. In particular, $\nicefrac{1}{2}\langle111\rangle$ loops are highly glissile, while their $\langle100\rangle$ counterparts are effectively immobile in most practical situations \cite{marian2003md}. As well, as indicated in \ref{app:energies}, the dissociation energies of point defects from each loop type can also differ substantially \cite{gilbert2008structure,mason2019relaxation}, which may have implications on their thermal stability during annealing.
However, being a mean-field approach, SCD does not discriminate between different Burgers vectors for the dislocation loops. Recent experiments using 800-keV Kr ions at 400$^\circ$C suggest a 50/50 partition between $\nicefrac{1}{2}\langle111\rangle$ and $\langle100\rangle$ Burgers vectors \cite{ZHENG2020162}. Similarly, TEM samples irradiated with 1.2-MeV W ions up to 0.017 dpa at room temperature indicate similar loop coexistence \cite{hasanzadeh2018three}.
Atomistic simulations generally predict a preponderance of $\nicefrac{1}{2}\langle111\rangle$ loops \cite{byggmastar2025four,alexander2016initio,mason2019relaxation}, although the observation of $\left<100\right>$ loops in MD simulations of high-energy cascades is not uncommon, and has been reported in some studies \cite{sand2013high,setyawan2015displacement,byggmastar2025four}.

\section{Conclusions}

We finish the paper with our most important conclusions and summary points.
\begin{itemize}
    \item We have assembled a number of fundamental physics features that have recently emerged in the literature into a numerical model for the accumulation of irradiation damage in tungsten. The model includes atomistically-generated Frenkel-pair insertion correlations and cluster size distributions, athermal SIA diffusion along rectilinear paths, and large sampling volumes to ensure that an adequate sampling of the high-energy tail of the PKA energy distribution is effectuated. 
    \item The model contains no adjustable parameters, and its accuracy is thus derived from the incorporation of  pre-computed atomistic properties and physics mechanisms primarily derived from detailed atomistic simulations.
    \item The model was validated through a series of XRDS experiments of self-ion-irradiated single-crystal W at room temperature. The experiments were designed to facilitate comparison with the model by simplifying the irradiation conditions and eliminating microstructural complexities.
    \item Model predictions are in excellent agreement with XRDS measurements of vacancy and SIA cluster size distributions at the end of a 0.008 dpa irradiation stage followed by a long annealing at room temperature. This agreement holds over a wide range of cluster sizes for both self-interstitial and vacancy defect types.
    \item The model is also in excellent agreement with other independent works on W irradiated under similar conditions near room temperature, predicting large densities of monovacancies and small vacancy clusters consistent with numerous PAS studies and predicting absolute defect cluster distributions in the larger TEM-visible size range comparable to existing studies.
    \item Finally, the model provides a comprehensive picture of the SIA and vacancy cluster size distributions in irradiated W, thus complementing the capabilities of disparate experimental techniques with differing sensitivities to specific cluster morphologies and sizes.
    
\end{itemize}

\section*{Acknowledgments}

The authors acknowledge financial support from the U.S.\ Department of Energy, Office of Fusion Energy Sciences (DOE-FES) under contracts DE-SC0018410 (S.H.~and J.M.) and DE-SC0022528 (B.S.~and G.T.).
This work used computational and storage services associated with the Hoffman2 Cluster which is operated by the UCLA Office of Advanced Research Computing’s Research Technology Group. 
The authors acknowledge the use of
facilities and instrumentation at the UC Irvine Materials
Research Institute (IMRI), which is supported in part by the
National Science Foundation through the UC Irvine Materials 
Research Science and Engineering Center (grant No.~DMR-
2011967).

\appendix
\section{Brief review of the stochastic cluster dynamics method}\label{app:scd}

The accumulation of defects during irradiation is simulated using the stochastic cluster dynamics (SCD) model. SCD is a stochastic variant of the mean-field rate theory (MFRT) technique, which relies on stochastic sampling of the underlying master equation for defect cluster evolution \cite{RN22,RN20}. Instead of deterministically solving exceedingly large sets of partial differential equations (PDEs) of the concentrations of defects (as in standard MFRT), SCD evolves an integer-valued defect population $N_i$ in a finite material volume $\Omega$, thus avoiding exponential growth in the number of ODEs. This makes SCD ideal to treat problems where the dimensionality of the cluster size space is high, e.g., when multispecies simulations are of interest \cite{DUNN201643,MARIAN2012293,hu2025helium}. 

SCD recasts the standard ODE system into stochastic equations of the form:
\begin{equation}\label{eq:scd}
\frac{dN_i}{dt}=\tilde{g}_i+\tilde{s}_{l}N_l- \left[\tilde{s}_{i}N_i+\sum_\alpha\tilde{s}_{\alpha i}N_i\right]+
\sum_{j}\left[\sum_{k}\tilde{K}_{jk}N_jN_k - \tilde{K}_{ij}N_jN_i\right]
\end{equation}
where the set $\{\tilde{g},\tilde{s},\tilde{K}\}$ represents the reaction rates for, respectively, defect insertion; thermal dissociation and annihilation at sinks; and binary reactions occurring inside $\Omega$. Subindices $j$ and $k$ refer to all distinct defect species present in the system, whereas the subindex $\alpha$ refers to the type of sinks available to absorb defects (`d' for dislocations, `gb' for grain boundaries, `ppt' for precipitates). The subindices in the expression satisfy $l-1=i$ and $j+k=i$ (reflecting dissociation by monomer emission and association of complementary clusters, respectively).
Equation \eqref{eq:scd} can be straightforwardly extended to partial differential equations to capture diffusion due to defect concentration gradients \cite{Yu_2020}.

Equation \eqref{eq:scd} is solved using the residence-time algorithm by sampling, selecting, and executing events from the set of rates $\{\tilde{g},\tilde{s},\tilde{K}\}$ \cite{RN22}. The volume $\Omega$ is arbitrary, although its minimum value is subjected to the numerical stability criterion:
\begin{equation}\label{eq:stability}
\Omega^{\frac{1}{3}}>\ell
\end{equation}
with
\begin{align}
\ell&=\max_i\{l_i\}\\
l_i&=\sqrt{D_i\tau_i}
\end{align}
where $D_i$ and $\tau_i$ are the diffusivity and the lifetime of mobile species $i$ with $\tau_i^{-1}=\tilde{s}+\sum_j\tilde{K}_{ij}N_j$. 
For consistency with the experimental conditions provided above, here we consider vacancies, self-interstitial atoms, and any cluster combination thereof. The number of clusters of type $i$ can be straightforwardly (and correctly) converted to a concentration by normalizing $N_i$ by the volume $\Omega$: $C_i=N_i/\Omega$.

As a practical note, under certain conditions the volume $\Omega$ can be rescaled for improved computational efficiency, which leads to an accelerated defect cluster evolution. A quantitative demonstration of volume rescaling applied to a selected group of defect clusters is provided in ref.~\cite{HE2024155325}. For more information about the model, the reader is referred to past publications from the authors \cite{RN22,RN20,Yu_2020}. 

\subsection{Defect insertion rates}

The defect insertion rates, $\tilde{g}_i$ in eq.\ \eqref{eq:scd}, are obtained for each defect type by multiplying the volumetric insertion rates ($g_i$ in Fig.\ \ref{fig:big-slide}) times the simulation volume $\Omega$. In turn, the $g_i$ rates are set by the ion insertion rate, defined in eq.\ \eqref{eq:dotg}, followed by sampling $C(E)$ recursively until the total ion energy expended on lattice damage, i.e., 2.65 MeV, is reached (see Sec.~\ref{sec:srim} for more details)\footnote{Note that using the ASTM-recommended value for the threshold displacement energy of W of $E_{\rm th}$=90 eV, this amount of energy would result in the creation of $N_{\rm NRT}=\eta E/2E_{\rm th}\approx2.35\times10^7$ Frenkel pairs according to the NRT formalism \cite{norgett1975proposed} when an efficiency of $\eta=0.8$ is used. However, this is known to grossly overpredict the number of defects actually created during irradiation \cite{zinkle2023quantifying}.}. 
During each sampling, defect distributions obeying eqs.\ \eqref{eq:corr} and using the scaling laws extracted from Fig.\ \ref{fig:barchart} are generated, and an array of species $i$ is created. These species are then added to the simulation volume and allowed to evolve according to eqs.\ \eqref{eq:scd}. 

In order to appropriately sample the entire range of cluster sizes illustrated by the histograms in Fig.\ \ref{fig:barchart}, it is desirable to produce PKA with energies greater than 100 keV, as, without them, the large clusters observed experimentally with $n>100$ would not be created in the simulations. However, due to the low probability of producing PKA with those energies (note that the average PKA energy in the present irradiations is $<$1 keV), large numbers of PKA samplings must be instantiated. This can only be achieved by maintaining sufficiently high values of $\dot G$ (eq.\ \eqref{eq:dotg}), which is directly proportional to the simulation volume $\Omega$. Thus, a crucial part of our model is ensuring that $\Omega$ is kept above a minimum limit that guarantees adequate sampling of $C(E)$ and the cluster size power laws.

\subsection{Defect energetics}\label{app:energies}

Defect mobilities display an Arrhenius temperature dependence, $D(T)=D_0\exp\left(-E_m/kT\right)$, where $D_0$ is the diffusion prefactor and $E_m$ is the migration energy. As indicated in Sec.\ \ref{sec:mob}, here we consider mobilities of single and di-interstitials, as well as monovacancies, with their disffusion parameters given in Table \ref{tab:mob}.

We also consider thermal dissociation of SIA and vacancy clusters by emission of monomers. As indicated in Fig.\ \ref{fig:big-slide}, the dissociation rate scales with the factor $\exp\left(-E^b_n/kT\right)$, where $E_n^b$ is the binding energy of a monomer (monovacancy or single SIA) to a cluster of size $n$. Atomistic calculations of SIA and vacancy clusters up to $n=8$ exist in the literature \cite{fikar2009molecular,becquart2010microstructural,alexander2016initio}. Specifically, here we use the DFT data by Becquart \etal~\cite{BECQUART201039}, as listed in Table \ref{tab:bind}. For larger sizes, the binding energies are obtained as:
\begin{equation}
    E^{b}_n= \left(E^{f}(1) + E^{f}(n-1)\right)-E^{f}(n)=E^{f}(1)-\left(E^{f}(n)-E^{f}(n-1)\right)\approx E^{f}(1)-\frac{dE^{f}(n)}{dn}
\label{eq:bind1}
\end{equation}
where $E^{{f}}(n)$ is the formation energy of a cluster of size $n$. 
Here we assume that vacancy clusters adopt a globular shape for $8\leq n<30$, whose formation energies are well captured using the capillary approximation, i.e., $E^{{f}}(n)\sim n^{\frac{2}{3}}$ \cite{becquart2010microstructural,Yu_2020}. The expression for the binding energies in this size range is given in Table \ref{tab:bind}. Finally, we assume that vacancy clusters with $n\geq30$ and SIA clusters with $n\geq8$ appear as prismatic loops with energies given by the expression \cite{alexander2016initio,mason2019relaxation}:
\begin{equation}
E^f(n)=a_0\sqrt{n}\log{n}+a_1\sqrt{n}+a_2
 \label{eq:bind2}
\end{equation}
Combined with eq.\ \eqref{eq:bind1}, we obtain the final expression for the binding energy of loops:
\begin{equation}
E^{b}_n= E^{f}(1)-\frac{a_0}{2\sqrt{n}}\log n+\frac{2a_0 + a_1}{2\sqrt{n}}=E^{f}(1)-\frac{a_0}{2\sqrt{n}}\left(\log n - 2- \frac{a_1}{a_0}\right)
 \label{eq:bind3}
\end{equation}
where $a_0$ and $a_1$ are numerical coefficients given for each type of loop in Table \ref{tab:bind}.
\begin{table}[ht!]
\centering
\caption{Binding energies of monomers to $I_n$ and $V_n$ clusters used in the simulations. 
All data from reference~\cite{becquart2010microstructural}. Values for $E^{f}(I_1)$ and $E^{f}(V_1)$  are 9.96 and 3.23~eV, respectively~\cite{ventelon2012ab}.}
\begin{tabular}{|l|c|}
\hline
Species & $E^b_n$ [eV] \\
\hline
$I_2$ & 2.12 \\
$I_3$ & 3.02 \\
$I_4$ & 3.60 \\
$I_5$ & 3.98 \\
$I_6$ & 4.27 \\
$I_7$ & 5.39 \\
$n > 7$ & $E^{f}(1)-\frac{a_0}{2\sqrt{n}}\log n+\frac{2a_0 + a_1}{2\sqrt{n}}$ \\
& $a_0=3.394$~~~$a_1=7.081$ \\
\hline
$V_2$ & $-0.10$ \\
$V_3$ & 0.04 \\
$V_4$ & 0.64 \\
$V_5$ & 0.72 \\
$V_6$ & 0.89 \\
$V_7$ & 0.72 \\
$7<n\leq30$ & $E^{f}({V}_1) + 1.71\left(E^{b}({V}_2) - E^{f}({V}_1)\right)\left[n^{2/3} - (n-1)^{2/3}\right]$ \\
$n > 30$ & $E^{f}(1)-\frac{a_0}{2\sqrt{n}}\log n+\frac{2a_0 + a_1}{2\sqrt{n}}$ \\
& $a_0=4.155$~~~$a_1=-0.01$ \\  
\hline
\end{tabular}
\label{tab:bind}
\end{table}



\newpage
\bibliography{IFE-refs,refs-he-h,refs-cascades,newrefs,XRDS}

\end{document}